\documentclass[sigconf]{acmart}
\usepackage[T1]{fontenc}
\usepackage{xcolor}
\usepackage{subfiles}

\usepackage[skip=6pt]{caption}

\PassOptionsToPackage{hyphens}{url}

\usepackage{graphicx}

\usepackage{booktabs}

\usepackage{amsmath}
\usepackage{algorithm}
\usepackage{algpseudocode}

\usepackage{adjustbox}
\newcommand{\headrow}[1]{\multicolumn{1}{c}{\adjustbox{angle=45,lap=\width-0.5em}{#1}}}

\newcommand{\full}{$\bullet$}
\newcommand{\prt}{$\circ$}

\makeatletter
\renewcommand\subsubsection{\@startsection
  {subsubsection}{3}{0mm}
  {-\baselineskip}
  {0.5\baselineskip}
  {\normalfont\large\bfseries}}
\makeatother

\usepackage{listings}
\newcommand*{\TitleFont}{%
      \usefont{\encodingdefault}{\rmdefault}{b}{n}%
      \fontsize{16}{20}%
      \selectfont}

\title{\vspace{-.5em}\TitleFont BlockSci: Design and applications of a blockchain analysis platform \vspace{0.5em}}

\author{Harry Kalodner}
\email{kalodner@cs.princeton.edu}
\affiliation{Princeton University}
\author{Steven Goldfeder}
\email{stevenag@cs.princeton.edu}
\affiliation{Princeton University}
\author{Alishah Chator}
\email{alishahc@cs.jhu.edu}
\affiliation{Johns Hopkins University}
\author{Malte M\"{o}ser}
\email{mmoeser@cs.princeton.edu}
\affiliation{Princeton University}
\author{Arvind Narayanan}
\email{arvindn@cs.princeton.edu}
\affiliation{Princeton University\vspace{1em}}

\setcopyright{none}
\renewcommand\footnotetextcopyrightpermission[1]{}
\begin{document}

\begin{abstract}
Analysis of blockchain data is useful for both scientific research and commercial applications. We present BlockSci, an open-source software platform for blockchain analysis. BlockSci is versatile in its support for different blockchains and analysis tasks. It incorporates an in-memory, analytical (rather than transactional) database, making it several hundred times faster than existing tools. We describe BlockSci's design and present four analyses that illustrate its capabilities.

This is a working paper that accompanies the first public release of BlockSci, available at \href{https://github.com/citp/BlockSci}{\UrlFont github.com/citp/BlockSci}. We seek input from the community to further develop the software and explore other potential applications.
\end{abstract}

\maketitle

\vspace{0.5em}

\section{Introduction}

Public blockchains constitute an unprecedented research corpus of financial transactions. Bitcoin's blockchain alone is 140 GB as of August 2017, and growing quickly. This data holds the key to measuring the privacy of cryptocurrencies in practice \cite{cookie-meets-blockchain,monero}, studying new kinds of markets that have emerged \cite{joinmarket,trends-tips-tolls}, and understanding the non-currency applications that use the blockchain as a database.

We present BlockSci, a software platform that enables the science of blockchains. It addresses three pain points of existing tools: poor performance, limited capabilities, and a cumbersome programming interface. BlockSci is 15x--600x faster than existing tools, comes bundled with analytic modules such as address clustering, exposes different blockchains through a common interface, imports exchange rate data and ``mempool'' data, and gives the programmer a choice of interfaces: a Jupyter notebook for intuitive exploration and C++ for performance-critical tasks.

BlockSci's design starts with the observation that blockchains are append-only databases; further, the snapshots used for research are static. Thus, the ACID properties of transactional databases are unnecessary. This makes an in-memory analytical database the natural choice. On top of the obvious speed gains of memory, we apply a number of tricks such as converting hash pointers to actual pointers, which further greatly increase speed and decrease the size of the data. We plan to scale vertically as blockchains grow, and we expect that this will be straightforward for the foreseeable future, as commodity cloud instances currently offer up to a {\em hundred times} more memory than required for loading and analyzing Bitcoin's blockchain. Avoiding distributed processing is further motivated by the fact that blockchain data is graph-structured, and thus hard to partition effectively. In fact, we conjecture that the use of a traditional, distributed transactional database for blockchain analysis has infinite COST \cite{mcsherry-cost}, in the sense that no level of parallelism can outperform an optimized single-threaded implementation.

BlockSci comes with batteries included. First, it is not limited to Bitcoin: a parsing step converts a variety of blockchains into a common, compact format. Currently supported blockchains include Bitcoin, Litecoin, Namecoin, and Zcash (Section \ref{sec:import}). Smart contract platforms such as Ethereum are outside our scope. Second, BlockSci includes a library of useful analytic and visualization tools, such as identifying special transactions (e.g., CoinJoin) and linking addresses to each other based on well-known heuristics (Section \ref{sec:analysis-library}). Third, we record transactions broadcast on the peer-to-peer network and expose them through the same interface. Similarly, we expose (historical and current) data on the exchange rates between cryptocurrencies and fiat currencies. These allow many types of analyses that wouldn't be possible with blockchain data alone.

The analyst begins exploring the blockchain through a Jupyter notebook interface (Section \ref{sec:interface}), which initially exposes a \texttt{chain} object, representing the entire blockchain. Startup is instantaneous because transaction objects are not initially instantiated, but only when accessed. Iterating over blocks and transactions is straightforward, as illustrated by the following query, which computes the average fee paid by transactions in each block mined in July 2017:

\begin{lstlisting}
fees = [mean(tx.fee() for tx in block) for 
        block in chain.range('Jul 2017')]
\end{lstlisting}

This interface is suitable for exploration, but for analyses requiring high performance, BlockSci also has a C++ interface. For many tasks, most of the code can be written in Python with a snippet of performance-sensitive code written as inline C++ (Section \ref{sec:interface}).

\begin{figure*}
\begin{centering}

  \includegraphics[width=0.8\linewidth]{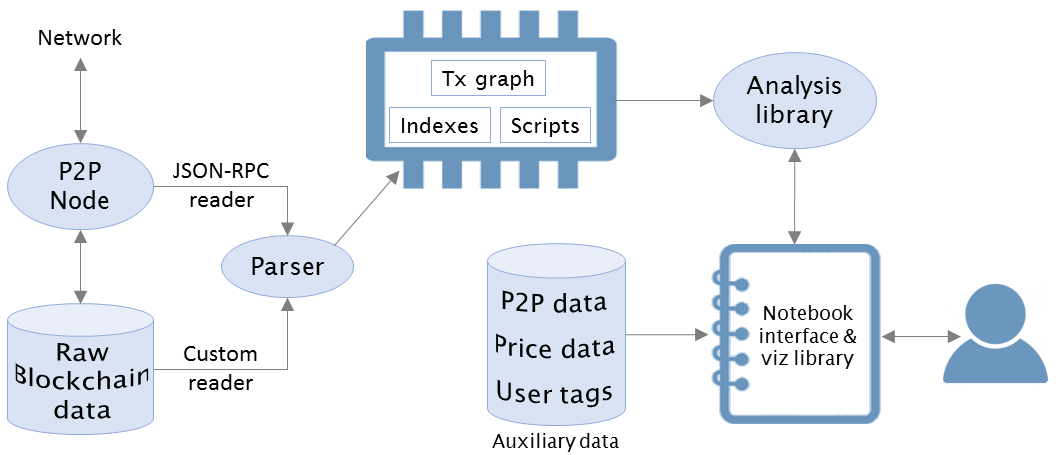}
  \caption{{\bf Overview of BlockSci's architecture.}}
    \label{fig:architecture}
  \end{centering}
\end{figure*}

In Section \ref{sec:applications} we present four applications to illustrate the capabilities of BlockSci. First, we show how multisignatures have the unfortunate effect of weakening confidentiality by exposing the details of access control on the blockchain, as suggested by Gennaro et al. \cite{ggn16}; multisignatures even hurt the privacy of users who {\em do not} use them (Section \ref{sec:multisig}). Next, we provide evidence that the cluster intersection attack  reported recently \cite{cookie-meets-blockchain} also works against Dash, a prominent privacy-focused altcoin with built-in mixing (Section \ref{sec:dash}). Turning to economics, we analyze the emerging market for block space, and identify behaviors by miners that result in foregoing significant transaction fees (\ref{sec:blockspace}). Finally, we provide improved estimates of the velocity of cryptocurrencies, i.e., the frequency with which coins change possession. This helps us understand their use as a store of value versus a medium of exchange.

\newpage

{\bf Overview.} Figure \ref{fig:architecture} shows an overview of BlockSci's architecture. There are two routes for importing data into BlockSci (Section \ref{sec:import}). Through either route, the data is converted into the same intermediate format for parsing (Section \ref{sec:parser}). The parser produces the Core Blockchain Data  (Section \ref{sec:dataformat}), which can be incrementally updated as new blocks come in. The analysis library (Section \ref{sec:analysis-library}) loads this data as an in-memory database, which the user can either query directly or through a Jupyter notebook interface (Section \ref{sec:interface}).

\subsection{Recording and importing data} 
\label{sec:import}
{\bf Supported blockchains.} Recall that the Bitcoin blockchain consists primarily of a directed acyclic graph of transactions. The edges connecting transactions have attributes, i.e., addresses or scripts, attached to them. Transactions are grouped into blocks which are arranged in a linear chain, with a small amount of metadata per block. BlockSci supports blockchains that follow this basic structure. For example, Litecoin makes no changes to the data structure, and is thus fully supported. 

Cryptocurrencies that introduce changes to the script operations may be supported only partially. Namecoin is supported, but the new script types it introduces are not parsed by BlockSci (the user can parse them with a few lines of code). Zcash is also supported, at least to the extent that Zcash blockchain analysis is even possible: it introduces a complex script that includes zero-knowledge proofs, but these aspects are parceled away in a special type of address that is not publicly legible by design.

An example of a currently unsupported blockchain is Monero because it doesn't follow the ``one-input, one-output'' paradigm. In other words, the transaction graph contains an additional type of node, the mixin. Supporting such blockchains would require changes to the internal logic as well as the programmer interface. Similarly, Ethereum departs from the transaction-graph model, and further, its script is vastly different from and more complex than Bitcoin's.
 
In our analyses we have worked with six blockchains: Bitcoin,\footnote{SegWit support is not yet included, but is planned shortly.} Bitcoin Cash, Litecoin, Namecoin, Dash, and ZCash. Many other cryptocurrencies make no changes to the blockchain format, and so should be supported with no changes to BlockSci.

{\bf Importer.} For altcoins with small blockchains where import performance is not a concern, we use the JSON-RPC interface that is supported by most altcoins. The advantage of this approach is versatility, as altcoins generally aim to conform to a standard JSON-RPC schema regardless of the on-disk data structures and serialization format. For larger blockchains (currently only Bitcoin is large enough for import performance to be a concern), we use our own high-performance importer that directly reads from the raw data on disk. Our Bitcoin importer also works on Litecoin and Dash as they use the same format.

The importer doesn't save data to disk; rather it passes data directly to the parser (Section \ref{sec:parser}), and the two execute in a pipelined fashion.

{\bf Mempool recorder.} BlockSci also records mempool data, that is, information about transactions that are broadcast to the P2P network and are waiting to be included in the blockchain. The waiting time of transactions provides valuable data about the block space market (and isn't recorded in the blockchain itself). Similarly, transactions that never make it into the blockchain are valuable for analysis. 

The mempool recorder has two modes. In {\em minimal mode}, it records only timestamps (equivalently, waiting times) of transactions that made it into the blockchain. Note that public sources of mempool data such as blockchain.info allow querying the timestamp by transaction hash, but not the bulk download of this data. In {\em full mode}, the recorder includes all information in the mempool, which encompasses transactions that were never included in a block. Timestamp data is loaded in memory for analysis whereas the full-mode data is stored on disk.

In any peer-to-peer system, different nodes will receive the same data at different times. Blockchain.info uses a geographically distributed set of nodes to obtain relatively accurate timestamps. BlockSci is a single-node system, so its timestamps inevitably lag those of blockchain.info. Based on 2 weeks of mempool data recorded by our AWS node in the \textsf{us-east-1d} data center, we found that our timestamps lag blockchain.info's timestamps by an average of 16 seconds and a standard deviation of 4 seconds. Any BlockSci user can perform a similar measurement and apply a uniform correction to eliminate the average lag, but of course the variance will remain.

\subsection{Parser}
\label{sec:parser}

The on-disk format of blockchains is highly inefficient for our purposes. It is optimized for a different set of goals such as validating transactions and ensuring immutability. Bitcoin Core and other such clients minimize memory consumption at the expense of disk space, whereas we aim for a single representation of the data that can fit in memory. A number of techniques help achieve this goal while simultaneously optimizing for speed of access:

\begin{enumerate}
\item Link outputs to the inputs that spend them in order to allow efficient graph traversal.
\item Replace hash pointers with IDs to shrink the data structure and optimize linkage.
\item Use fixed size encodings for data fields whenever possible.
\item De-duplicate address/script data.
\item Optimize the memory layout for locality of reference. 
\end{enumerate}

\begin{table}[t]
\begin{minipage}{.45\linewidth}
\centering
\vspace{-2.45em}
\begin{tabular}{l r}
\toprule
Description    & Size    \\ \midrule
Spent/spending tx ID & 32 bits \\ 
Address ID     & 32 bits \\
Value   & 60 bits \\
Address Type   & \phantom{0}4 bits \\
\bottomrule
\end{tabular}
\caption{Input/output structure}
\label{tab:inout}
\end{minipage}\hfill
\begin{minipage}{.45\linewidth}
\centering
\begin{tabular}{l r}
\toprule
Description    & Size    \\ \midrule
Size & 32 bits \\
Locktime & 32 bits \\
Input count     & 16 bits \\
Output count   & 16 bits \\
Outputs   & 128 bits each \\
Inputs   & 128 bits each \\
\bottomrule
\end{tabular}
\caption{Transaction structure}
\label{tab:transaction}
\end{minipage}
\end{table}

{\bf Parsing is sequential and stateful.} 
The blockchain must be processed sequentially because two types of state are required to transform the blockchain into the BlockSci analysis format. Each transaction input specifies which output it spends, encoded as (transaction hash, output index). To transform the transaction hash into the ID that BlockSci assigns to the transaction, the parser must maintain the hash $\rightarrow$ ID map. Similarly, it must maintain a mapping from addresses to IDs for linking and deduplication.

The transaction hash $\rightarrow$ ID map can be made smaller by pruning transaction hashes for which all the outputs of the transaction have been spent. Address mapping, however, allows no such optimization. Any address may be used by any output and thus all addresses must be tracked at all times. Storing the map in memory would require too much memory, and storing it on disk would make the parser too slow.

\textbf{Optimization: LRU Cache and Bloom filter.}
To achieve further optimizations, we observe that the vast majority of inputs spend recently created outputs (e.g., 89\% of inputs spend outputs created in the last 4000 blocks). Similarly, the vast majority of addresses that are ever used again are used soon after their initial usage (e.g., 90\% within 4000 blocks).

This allows the following trade-off between speed and memory consumption:
\begin{itemize}
\item The transaction and addresses hashes are stored in a key-value database on disk (LevelDB), with a memory cache that has a Least Recently Used replacement policy. The cache also contains (and does not evict) all addresses that have been used multiple times, which is a  small fraction of addresses (6.8\%).
\item A bloom filter stores the list of seen addresses. If an address is not in the cache, the bloom filter is queried before the database. Recall that negative results from a bloom filter are always correct, whereas there is a small chance of false positives. This ensures correctness of the lookup while minimizing the number of database queries for nonexistent addresses.
\end{itemize}

Another optimization is that since the parser takes as input the serialized blockchain, we assume that transactions and blocks have been validated by the peer-to-peer node before being saved. This allows us to forgo the vast majority of script processing.

\textbf{Incremental Updates}: The append-only nature of the blockchain enables incremental updates to the parser output. The parser serializes its final state at the end of a run and resumes from that state when invoked again. The main difficulty with this approach is handling blockchain reorganization which occurs when a block that was originally in the longest branch is surpassed by a different branch. This requires reversing the parser process on the previous blocks before applying the new ones (see also the discussion of the snapshot illusion in Section \ref{sec:analysis-library}).

\subsection{Core Blockchain Data}
\label{sec:dataformat}

The output of the parser is the Core Blockchain data, which is the primary dataset for analysis.

{\bf Transaction graph.} The transaction graph is stored in a single sequential table of transactions, with entries having the structure shown in Table \ref{tab:transaction}. Note that entries have variable lengths, due to the variable number of inputs and outputs (there is a separate array of offsets for indexing, due to the variable entry lengths). Normally this would necessitate entries to be allocated in the heap, rather than contiguously, which would have worse memory consumption and worse locality of reference.

However, because of the append-only property of the blockchain, there are only two types of modifications that are made to the transactions table: appending entries (due to new transactions) and length-preserving edits to existing entries (when existing outputs are consumed by new transactions). This allows us to create a table that is stored as flat file on disk that grows linearly as new blocks are created. To load the file for analysis, it is mapped into memory. The on-disk representation continues to grow (and be modified in place), but the analysis library provides a static view (Section \ref{sec:analysis-library}).

{\bf Layout and locality.} The main advantage of the transaction graph layout is spatial locality of reference. Analyses that iterate over transactions block-by-block exhibit strong locality and benefit from caching. Such analyses will remain feasible even on machines with insufficient memory to load the entire transaction graph, because disk access will be sequential.

The layout stores both inputs and outputs as part of a transaction, resulting in a small amount of duplication (a space cost of about 19\%), but resulting in a roughly 10x speedup for sequential iteration compared to a normalized layout.
Variants of the layout are possible depending on the types of iteration for which we wish to optimize performance (Section \ref{sec:performance}). 

{\bf Indexes.} The transaction graph data structure does not include transaction hashes or addresses. The mapping from transaction/address IDs to hashes (and vice versa) is stored in separate indexes. Accessing these indexes is almost never performance critical in scientific analysis --- in fact, many analyses don't require the indexes at all. Due to the size of the files (25 GB for the transaction index and 29 GB for the address index for the current Bitcoin blockchain), users may not want to load them in memory for analyses where they are not performance critical. Thus, we store them in a SQLite database. SQLite has a command-line parameter that allows configuring the amount of memory used for caching. Currently these are the only indexes in BlockSci. Other indexes on attributes such as transaction fees are planned for the future.

{\bf Scripts.}  BlockSci currently categorizes scripts into 5 types: pay-to-public-key-hash, pay-to-script-hash, multisig, pubkey, and null data (OP\_RETURN). All other scripts are categorized as nonstandard. We plan to add support for more script types, including those found in altcoins but not Bitcoin. For scripts belonging to any of the supported types, BlockSci parses the script and stores information relevant to analysis, while discarding unnecessary script data. For pubkey and pay-to-public-key-hash this means that we record the pubkeyhash and pubkey when available (i.e., if the output has been spent). For pay-to-script-hash we record the script hash as well as a reference to the address it contains (recursively one of  the types defined above). For multisig we record pointers to the pubkey addresses that can spend the multisig as well as the number of addresses required to spend it. For null data we record the data store. For nonstandard types we record the entire script, allowing the user to write their own parsing code as necessary.

\subsection{BlockSci Analysis Library}
\label{sec:analysis-library}

{\bf Memory mapping and parallelism.} Since BlockSci uses the same format for the transaction graph on disk and in memory, loading the blockchain simply involves memory-mapping this file. Once in memory, each transaction in the table can be accessed as a C++ {\tt struct}; no new memory needs to be allocated to enable an objected-oriented interface to the data.

Another benefit of memory mapping is that it allows parallel processing with no additional effort, via a multithreaded or multiprocess architecture. Recall that if a file is mapped into memory by multiple processes, they use the same physical memory for the file. The file has only one writer (the parser); it is not modified by the analysis library. Thus, synchronization between different analysis instances isn't necessary.  With a disk-based database, analyses tend to be I/O-bound, with little or no benefit from multiple CPUs, whereas BlockSci is CPU-bound, and speed is proportional to the number of CPUs used (Section \ref{sec:performance}). Memory mapping also makes it straightforward to support multiple users on a single machine, which is especially useful given that Jupyter notebook (the main interface to BlockSci) can be exposed via the web. 

{\bf The snapshot illusion.} The following three seemingly contradictory properties hold in BlockSci:
\begin{enumerate}
\item The transactions table is constantly updated on disk as new blocks are received (note that arbitrarily old transactions may be updated if they have unspent outputs that get spent in new blocks)
\item The table is memory-mapped and shared between all running instances of BlockSci
\item Each instance loads a snapshot of the blockchain that never changes unless the programmer explicitly invokes a reload.
\end{enumerate}

The contradiction disappears once we notice that the state of the transactions table at any past point in time (block height) can be reconstructed given the current state. To provide the illusion of a static data structure, when the blockchain object is initialized, the {\tt maxHeight} attribute stores the height of the blockchain at initialization time. The blockchain height on disk increases over time, but the {\tt maxHeight} attribute remains fixed, and accesses to blocks past this height are not possible. The analysis library intercepts accesses to transaction outputs, and rewrites them so that outputs that were spent in blocks after {\tt maxHeight} are treated as unspent.

BlockSci currently exposes only the longest chain and hides orphaned/stale blocks. The library seeks to ensure that when the user reloads the chain, it will be a superset of the previous snapshot; in other words, it aims to hide reorganizations (reorgs) of the blockchain. This is done by ignoring the most recent few blocks during initialization. The probability of a reorg that affects $d$ or more blocks decreases exponentially in $d$. The default value of $d$ is 6. If a deeper reorg happens, the analysis library throws an exception.

{\bf Mapreduce.} Many analysis tasks, such as computing the average transaction fee over time, can be expressed as mapreduce operations over the transactions table (or ranges of blocks). Thus the analysis library supports a mapreduce abstraction. An additional advantage is parallelism: with no additional effort from the programmer, the library handles parallelizing the task to utilize all available cores. As we show in Section \ref{sec:basic-runtime-stats}, iterating over all transactions, transaction inputs, and transaction outputs on the Bitcoin blockchain as of August 2017 takes only 10.3 seconds on a {\em single} 4-core EC2 instance.

{\bf Address linking.} Recall that cryptocurrency users can trivially generate new addresses, and most wallets take advantage of this ability. Nevertheless, addresses controlled by the same user or entity may be linked to each other, albeit imperfectly, through various heuristics. Address linking is a key step in analytic tasks including understanding trends over time and evaluating privacy.

Meiklejohn et al. proposed two address-linking heuristics \cite{meiklejohn2013fistful}: (1) inputs spent to the same transaction are controlled by the same entity and (2) change addresses are not reused. We add an exception to heuristic 1: it isn't applicable to CoinJoin transactions. This requires accurately detecting CoinJoin transactions; we use the algorithm described in Goldfeder at al. \cite{cookie-meets-blockchain}.

These heuristics create links (edges) in a graph of addresses. By iterating over all transactions and applying the union-find algorithm on the address graph, we can generate clusters of addresses. This set of clusters is the output of address linking. We use the union-find implementation by Jakob \cite{union-find}. 

\begin{figure}
\centering
\includegraphics[width=\columnwidth]{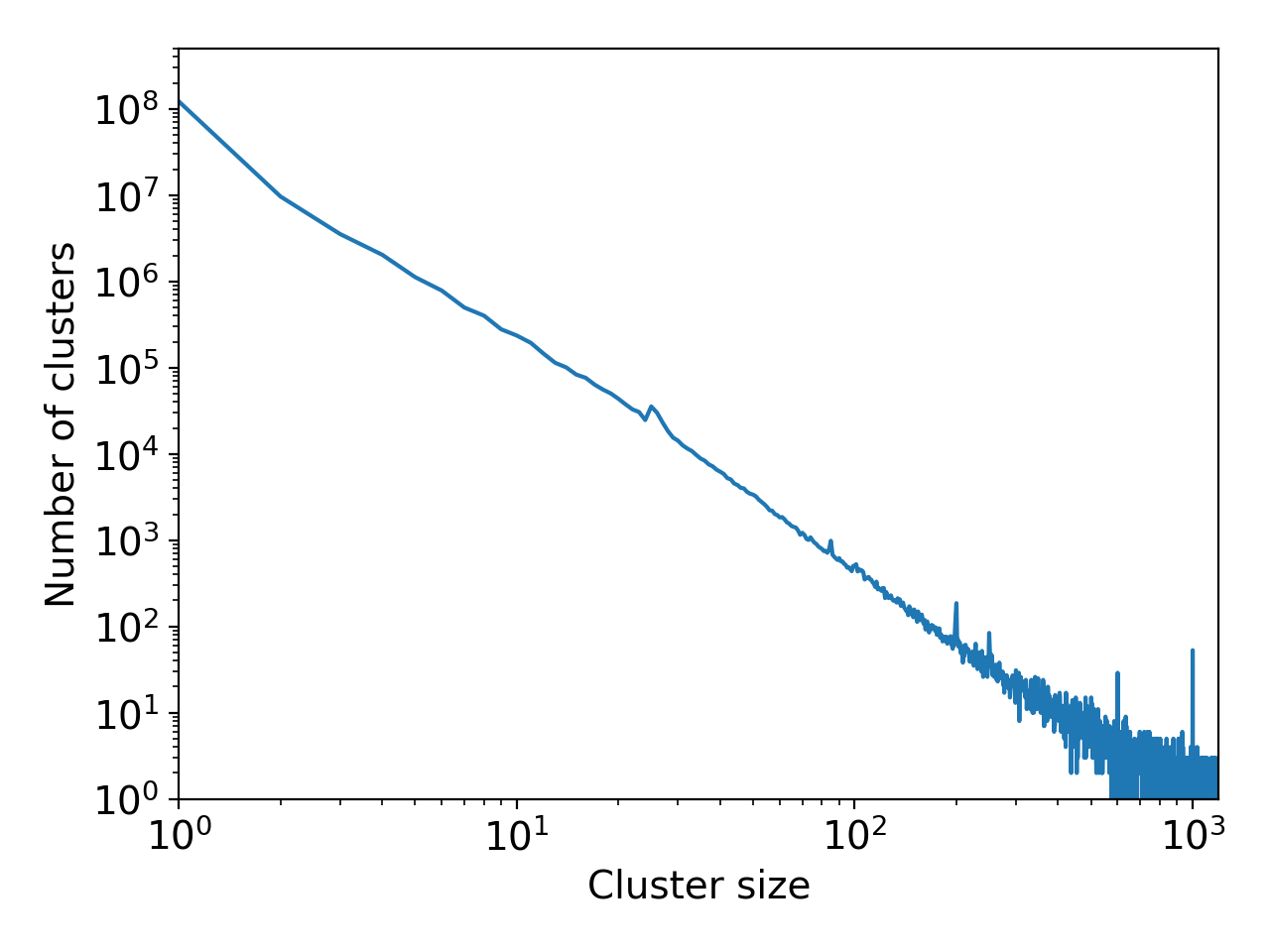}
\caption{Distribution of sizes of address clusters in Bitcoin after applying address-linking heuristics. Sizes 1--2,000 are shown here but there are many clusters that are much larger.}\label{fig:cluster-size-distribution}
\vspace{-1em}
\end{figure}

Figure \ref{fig:cluster-size-distribution} shows the distribution of cluster sizes. There are about 145 million clusters in total, of which about 122 million are single addresses, and about 20 million have between 2 and 20,000 addresses. There are 13 clusters with over 20,000 addresses, including one supercluster with over 139 million addresses.

Address linking is inherently imperfect, and ground truth is difficult to obtain on a large scale, since it requires interacting with service providers. Many other heuristics are possible, including those that account for the behavior of specific wallets. We do not attempt to be comprehensive, resulting in false negatives (i.e., missed edges, resulting in more clusters than truly exist). More perniciously, most of the heuristics are also subject to false negatives (i.e., spurious edges), which can lead to ``cluster collapse''. In particular, it is likely that the supercluster above is a result of such a collapse.

Considering the evolving nature of address linking techniques, and considering that different sets of heuristics may be suited to different applications, we provide an easy way for the programmer to recompute address clusters using their own set of heuristics. We conjecture that spectral clustering techniques \cite{ng2002spectral} can minimize false positives and negatives and largely obviate the need for tediously compiled manual heuristics. This is a topic for future work.

{\bf Tagging.} Address linking is especially powerful when combined with address tagging, i.e., labeling addresses with real-world identities. This can be useful for forensics and law-enforcement investigations but it can also violate user privacy. BlockSci does not provide address tags. Tagging requires interacting with service providers and cannot be done in an automated way on a large scale. Companies such as Chainalysis and Elliptic specialize in tagging and forensics, and blockchain.info allows users to publicly tag addresses that they control. BlockSci has a limited tagging feature: if the user provides tags for a subset of addresses, the address-linking algorithm will propagate those tags during the address linking step.

\subsection{Programmer interface}
\label{sec:interface}

Jupyter notebook is a popular Python interface for data science. It allows packaging together code, visualization, and documentation, enabling easy sharing and reproducibility of scientific findings. We expose the C++ BlockSci library to Python through the Pybind11 interface. While we intend Jupyter notebook to be the main interface to BlockSci, it is straightforward to utilize the analysis library directly from standalone C++ or Python programs and derive most of the benefits of BlockSci. Bindings for other languages may be added in the future.

Python is not a language known for performance; unsurprisingly, we find that it is significantly slower to run queries through the Python interface. Nevertheless, our goal is to allow the programmer to spend most of their time interacting with the Jupyter notebook, while simultaneously ensuring that the bottleneck parts of queries execute as C++ code. This is a difficult tradeoff, and is a work in progress. We illustrate this through an example. 

Suppose our goal is to find transactions with anomalously high transaction fees --- say 0.1 bitcoins ($10^7$ satoshis), worth several hundred US dollars at the time of writing. The slowest way to do this would be to write the entire query in Python:

\begin{figure}
\centering
\includegraphics[width=\columnwidth]{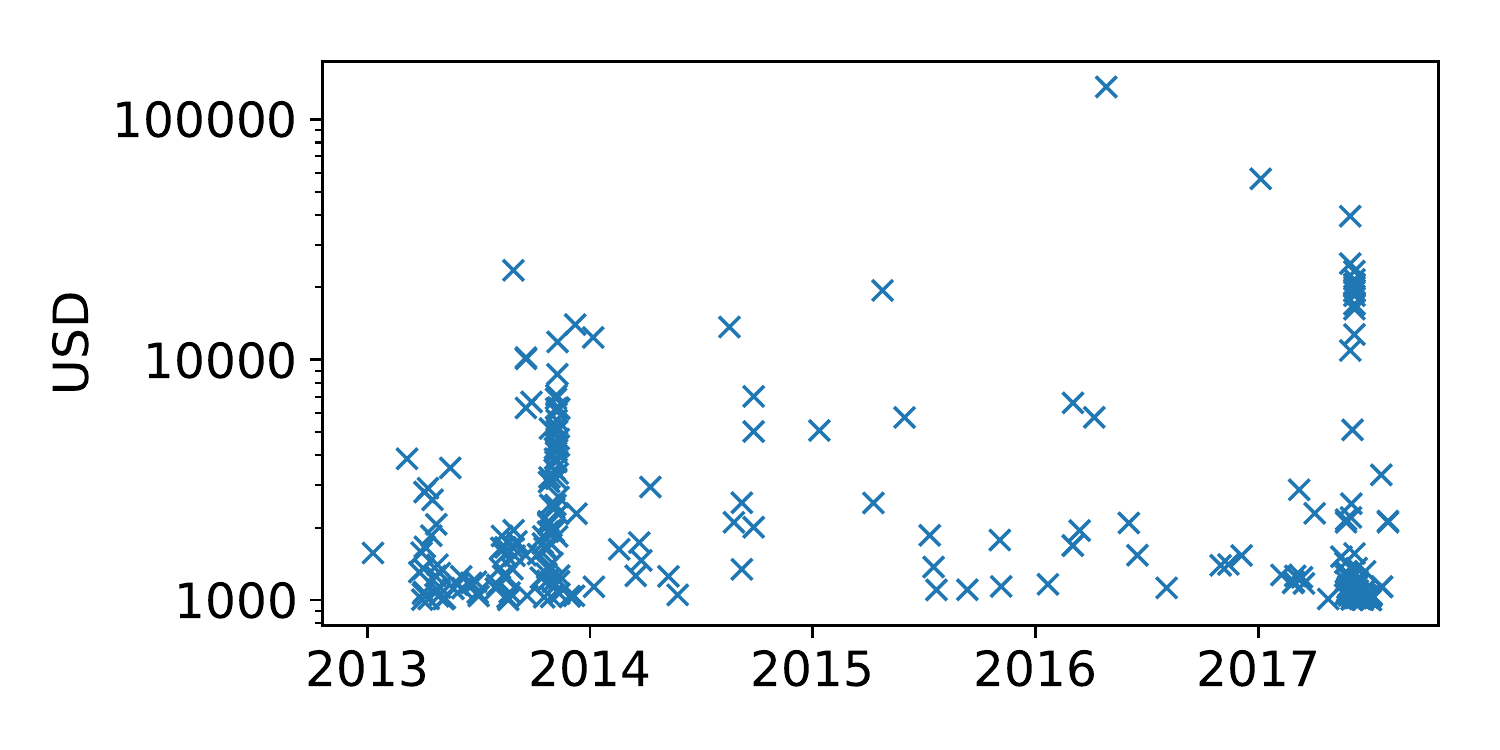}
\caption{Bitcoin transactions with fees worth over USD 1,000 at the time of the transaction. Note the log scale.}\label{fig:high-fee}
\end{figure}

\begin{lstlisting}
[tx for block in chain for tx in block if sum(txin.value for txin in tx.txins) - sum(txout.value for txout in tx.txouts) > 1e7]
\end{lstlisting}

This way does not result in acceptable performance. However, there is a simple way to improve both performance and conciseness:

\begin{lstlisting}
[tx for block in chain for tx in block if tx.fee() > 1e7]
\end{lstlisting}

A variant of this syntax automatically enables multithreading: 

\begin{lstlisting}
chain.filter_tx(lambda tx: tx.fee() > 1e7)
\end{lstlisting}

{\tt tx.fee()} is just one of many helper functions exposed by the Python library that execute as C++. Another such function is \\
{\tt block.total\_out()} which returns the total output value of transactions in the block. We've found that most of the analyses discussed in Section \ref{sec:applications} can benefit from a small number of such helper functions.

Another common paradigm is a selector. A small snippet of inline C++ code can be invoked through the notebook to return a subset of transactions (support for a declarative syntax rather than C++ code is planned for the near future). This subset of interest can then be processed in Python. The selector paradigm is a good fit for the anomalous-fee query:

\begin{lstlisting}
chain.cpp.filter_tx("tx.fee() > 10000000")
\end{lstlisting}

\begin{table}[]
\centering
\begin{tabular}{l r r}
\toprule
Iterating over        & Single Threaded & Multithreaded \\ \midrule
Transaction headers & 13.1 sec         & 3.2 sec      \\ 
Transaction outputs & 27.9 sec        & 6.6 sec      \\ 
Transaction inputs \& outputs  & 46.4 sec        & 10.3 sec      \\ 
Headers in random order        & 303.0 sec        & Unsupported              \\
\bottomrule
\end{tabular}
\caption{BlockSci C++ running time for various queries iterating over 478,449 blocks.}
\label{tab:basic-runtime-stats}
\end{table}

Here {\tt chain.cpp} encapsulates a set of functions that pass C++ code to the analysis library. This is the fastest way to write this query from the Python interface. We provide performance figures for all the above syntaxes in Section \ref{sec:basic-runtime-stats}.

Incidentally, the highest transaction fee that has ever been paid is 291 BTC. On April 26, 2016, the creator of a transaction famously and accidentally swapped the value and the fee, losing the equivalent of USD 136,000 at the time. In fact, there are 300 transactions with a fee over 1000 USD. We visualize these in Figure \ref{fig:high-fee}.

\subsection{Performance evaluation}
\label{sec:performance}

We now report the speed and memory consumption of BlockSci. A few notes on the setup:
\begin{itemize}
\item All measurements were performed on a single EC2 instance  (8 vCPUs, 2.5 GHz, Intel Xeon E5-2670v2, 61 GiB memory, 1 x 160 GiB Storage Capacity). The cost is 66 US cents per hour.
\item All measurements assume that the in-memory data structures are already loaded in memory. This takes about 60 seconds and needs to be done only once per boot.
\item By default all measurements are for the C++ interface; we report the performance of the Python interface separately.
\item By default all measurements are performed on the Bitcoin blockchain as of August 2017 (block count 478,559).
\end{itemize}

\subsubsection{Basic run time statistics}
\label{sec:basic-runtime-stats}

The most common type of access is a mapreduce-style iteration over the blockchain. A representative example is finding transactions with anomalously high fees, because computing the fee requires iterating over not just transactions, but also the inputs and outputs of each transaction. In essence, this query touches the entirety of the transactions table data. As Table \ref{tab:basic-runtime-stats} shows, a {\em single-threaded} implementation of this query completes in 46 seconds. Mapreduce-style queries are embarrassingly parallel, as seen in the table. Our test machine has 8 virtual cores, i.e., 4 physical cores with hyperthreading. The maximum possible speedup achievable is slightly over 4x, and this speedup is achieved.

The table shows that iterating over only the outputs (e.g., finding the max output value) is faster, and iterating over only the headers (e.g., finding transactions with a given value of nLockTime) is faster still. 

The above queries benefit from locality of reference. Other queries, especially those involving graph traversal, will not. To simulate this, we recomputed the query that examines transaction headers, this time iterating over the transactions in random order. We see that there is a 23-fold slowdown.

In Section \ref{sec:interface} we presented several paradigms for querying the blockchain from the Python interface: pure Python, C++ helper functions, and C++ selector. Figure \ref{tab:python-runtime-stats} shows the performance of these three paradigms on the anomalous-fee query. We see that the pure-Python method has unacceptable performance, the helper method is faster but still slow, and the C++ selector method is (unsurprisingly) essentially as fast as running the query in C++.

\begin{table}[]
\centering
\begin{tabular}{l r r}
\toprule
Query type         & Single threaded & Multithreaded \\ \midrule
Pure python              & 11 hrs       & 2.8 hrs       \\ 
Using C++ builtin        & 32 min       & 14 min       \\ 
Using C++ selector           & 47 sec       & 11.4 sec       \\ 
\bottomrule
\end{tabular}
\caption{BlockSci Python running time for the anomalous-fee query iterating over 478,559 blocks under the three paradigms discussed in Section \ref{sec:interface}.}
\label{tab:python-runtime-stats}
\end{table}

\subsubsection{Comparison with previous tools}
\label{sec:performance-comparison}

In comparing BlockSci with previous tools (some of which are special-purpose blockchain analysis tools, and others are databases that have been used for blockchain analysis) we have attempted to make the comparisons as fair as possible. We have used the same hardware when possible, and we always use benchmark tests that were used by the authors of the respective tools. A perfectly fair comparison may not always be possible; the main import of this section is that BlockSci is generally orders of magnitude faster than these tools.

Rubin presents BTCSpark \cite{btc-spark}, a distributed blockchain analysis platform based on Apache Spark. A performance benchmark reported in the paper is the ``TOAD'' query, for Total Output Amount Distribution. With 10 EC2 instances, all m3.large (6.5 ECUs, 2 vCPUs, 2.5 GHz, Intel Xeon E5-2670v2, 7.5 GiB memory, 1 x 32 GiB Storage), BTCSpark takes 3.7 minutes to execute TOAD on a block count of around 390,000. On our test EC2 instance, BlockSci executes  this query in 28.3 seconds. The dollar cost of this query is 15x lower for BlockSci than for BTCSpark with this configuration. The run time of BTCSpark appears to taper off at around 10 instances; thus, BlockSci on a single instance is likely significantly faster than BTCSpark with any number of instances.

M{\"o}ser and B{\"o}hme used the Neo4j graph database for processing the Bitcoin blockchain \cite{trends-tips-tolls, joinmarket}. We obtained their Neo4j database (which included blocks up to height 419,094) and instantiated it on our test instance. Neo4j supports a declarative graph query language, Cypher, as well as a Java API that compiles to low-level code. We implemented all three of the analyses reported in Table \ref{tab:basic-runtime-stats} via the Java API, as it is significantly faster. We obtained running times of 53 seconds, 2,300 seconds and 3,700 seconds respectively for the three queries. Since the Neo4j implementation is I/O-bound, parallelization on a single instance isn't possible. For the same block height,  BlockSci executes these queries in 2.0 seconds, 3.9 seconds, and 6.0 seconds respectively in multithreaded mode. Thus, on a single instance, BlockSci is 27x--600x faster.

The parsing tool BlockParser \cite{blockparser} is often used as an analysis tool as well, and explicitly supports this functionality by providing hooks for the programmer to insert analysis code that can be called while parsing. It comes with the ``Simple Stats'' benchmark (computing average input count, average output count, and average value). On our test instance, Blockparser takes 1,190 seconds to execute this query. BlockParser is single-threaded, and would be difficulty to parallelize due to the statefulness of parsing. With BlockSci, the single-threaded implementation runs in 30.9 seconds and the multithreaded implementation in 9.1 seconds, a 39x--131x speedup.

Finally, Bartoletti et al. present a Scala-based blockchain analysis library \cite{bartoletti-general}. A direct performance comparison is difficult, since their framework requires a time-consuming step to create queries (requiring up to tens of hours), followed by a faster query execution step. Of their 5 benchmarks, the {\em fastest} query (``OP\_RETURN metadata'') requires 2 hours to create and 0.5 seconds to execute. BlockSci executes this in 7.5 seconds, slower than their query execution time but 960x faster than their query creation time. Another query, ``transaction fees'', requires a creation time of 35 hours and executes in 448 seconds. BlockSci completes this query in 30.2 seconds, 4172x faster than their query execution time and 14x faster than their query creation time. 

Bartoletti et al. carry out their experiments on a PC with a quad-core Intel Core i5-4440 CPU @ 3.10GHz, equipped with 32GB of RAM and 2TB of hard disk storage. This is less memory than our test instance, but a more powerful CPU and far more storage.

We note that while workloads such as blockchain statistics sites (e.g., \url{https://blockchain.info/charts}) might consist of running the same set of queries at regular intervals, scientific workloads are characterized by a diversity of queries, and hence effective research tools must avoid large creation times for new queries.

\begin{table}[t]
\vspace{4.5em}
\centering
\begin{tabular}{l c c}
\toprule
           & Growth (bytes)           & Current    \\ \midrule
Current    & 20\,$N_{tx}$ + 16\,$N_{in}$ + 16\,$N_{out}$ & 25.21 GB \\
Normalized & 20\,$N_{tx}$ + \phantom{0}8\,$N_{in}$ + 16\,$N_{out}$  & 20.34 GB \\
64-bit     & 20\,$N_{tx}$ + 24\,$N_{in}$ + 24\,$N_{out}$ & 35.39 GB \\
Fee Cached & 30\,$N_{tx}$ + 16\,$N_{in}$ + 16\,$N_{out}$ & 27.6 GB \\
\bottomrule
\end{tabular}
\caption{Size of the transaction graph under each of 4 possible memory layouts. The `Current' column refers to the Bitcoin blockchain as of the end of July 2017, which has about 243 million (nodes) transactions and 663 million edges (outputs, including unspent ones).}
\label{tab:memory}
\vspace{-1.5em}
\end{table}

\subsubsection{Parser performance}
\label{sec:parser-performance}

Parsing the blockchain needs to be done only once upon installation; incremental updates are essentially instantaneous. We configured the parser with an 8 GB cache; this resulted in a run time of 11 hours. Faster performance is possible with a larger cache. Note that Bitcoin Core takes several hours to download the blockchain, so initialization is slow anyway. In the future we plan to distribute the Core Blockchain Data (serialized using Protocol Buffers) with regular incremental updates, so that BlockSci users can avoid a time-consuming initialization step, or even having to run a P2P node at all, unless the analysis task requires mempool data.

\subsubsection{Memory}
\label{sec:memory}

Table \ref{tab:memory} shows the memory consumption of BlockSci as a function of the size of the blockchain (measured by the number of transactions, inputs, outputs, and addresses). As noted earlier, for all analysis tasks we have encountered so far, only the transaction table needs 
to be in memory to ensure optimal performance. As of August 2017, this comes out to 22 GB for Bitcoin.

Recall that BlockSci's default layout of the transaction table is not normalized: coins are stored once as inputs and once as outputs. The table also shows the memory consumption for several alternate layouts. Although normalizing the layout would save 21\% space, it leads to a steep drop in performance for typical queries such as max-fee. Alternatively, we could  store derived data about transactions, such as the fee, at the expense of space. Finally, we also show how the space consumption would increase if and when we need to transition to 64-bit integers for storing transaction and address IDs.

\section{Applications}
\label{sec:applications}

We now present four analyses that highlight BlockSci's effectiveness at supporting blockchain science. The first two relate to privacy and confidentiality, and the latter two relate to the economics of cryptocurrencies. Table \ref{tab:blocksci-features} shows how these applications take advantage of the features of BlockSci's analysis library and data sources.

\begin{table}
\centering
\label{tab:systemization}
\begin{tabular}{l cccccc }
\multicolumn{1}{l}{Application}    &\headrow{Mapreduce queries} &\headrow{Address linkage}   &\headrow{Script parsing}   &\headrow{Mempool data}    &\headrow{Exchange rate data}&\headrow{Altcoin support}\\\midrule
Multisig (Sec. \ref{sec:multisig})	&\full &\full &\full &{} &{} &{} \\
Dash privacy (Sec. \ref{sec:dash})	    	&\full &\prt &{} &{} &{} &\full		\\
Block space (Sec. \ref{sec:blockspace}) &\full&{}	&{}	&\full	&\full&{}\\
Velocity (Sec. \ref{sec:velocity})	&\full		&\full &{}	&{}	&\full	&\full	\\
 \bottomrule
\end{tabular}
\caption{Usage of BlockSci features and data sources in various analyses. Note: the address-linkage algorithm needed to be reimplemented for Dash due to differences in transaction structure.}
\label{tab:blocksci-features}
\end{table}

\subsection{Multisignatures hurt confidentiality}
\label{sec:multisig}
Security conscious users or companies that store large amounts of cryptocurrency often make use of Bitcoin's multisignature capability. Unlike standard pay-to-public-key-hash (P2PKH) transactions which only require one signature to sign, multisig addresses allow one to specify $n$ keys and a parameter $m \leq n$ such that $m$ of the specified keys need to sign in order to spend the money. This feature makes it possible to distribute control of a Bitcoin wallet: keys can be stored on $n$ servers or by $n$ different employees of a company such that $m$ of them must agree to authorize a transaction. A typical example of this would be for a user to keep a key on both her desktop computer and her smartphone and require the participation of both to authorize a transaction (a 2-out-of-2 multisig). Almost always with multisig scripts, pay-to-script-hash (P2SH) transactions are used, which is a transaction type in which the address to which the money is sent is a hash of the redeem script. As of August 2017, about 13\% of all bitcoins are held in multisig addresses.

In this section we show how multisignatures expose confidential information about access control on the blockchain, as suggested by Gennaro et al~\cite{ggn16}. We further show how the use of multisignatures can hurt the privacy of {\em other} users. Finally, we find patterns of multisig usage that substantially reduce its security benefits.

\begin{figure}
\centering
\includegraphics[width=\columnwidth]{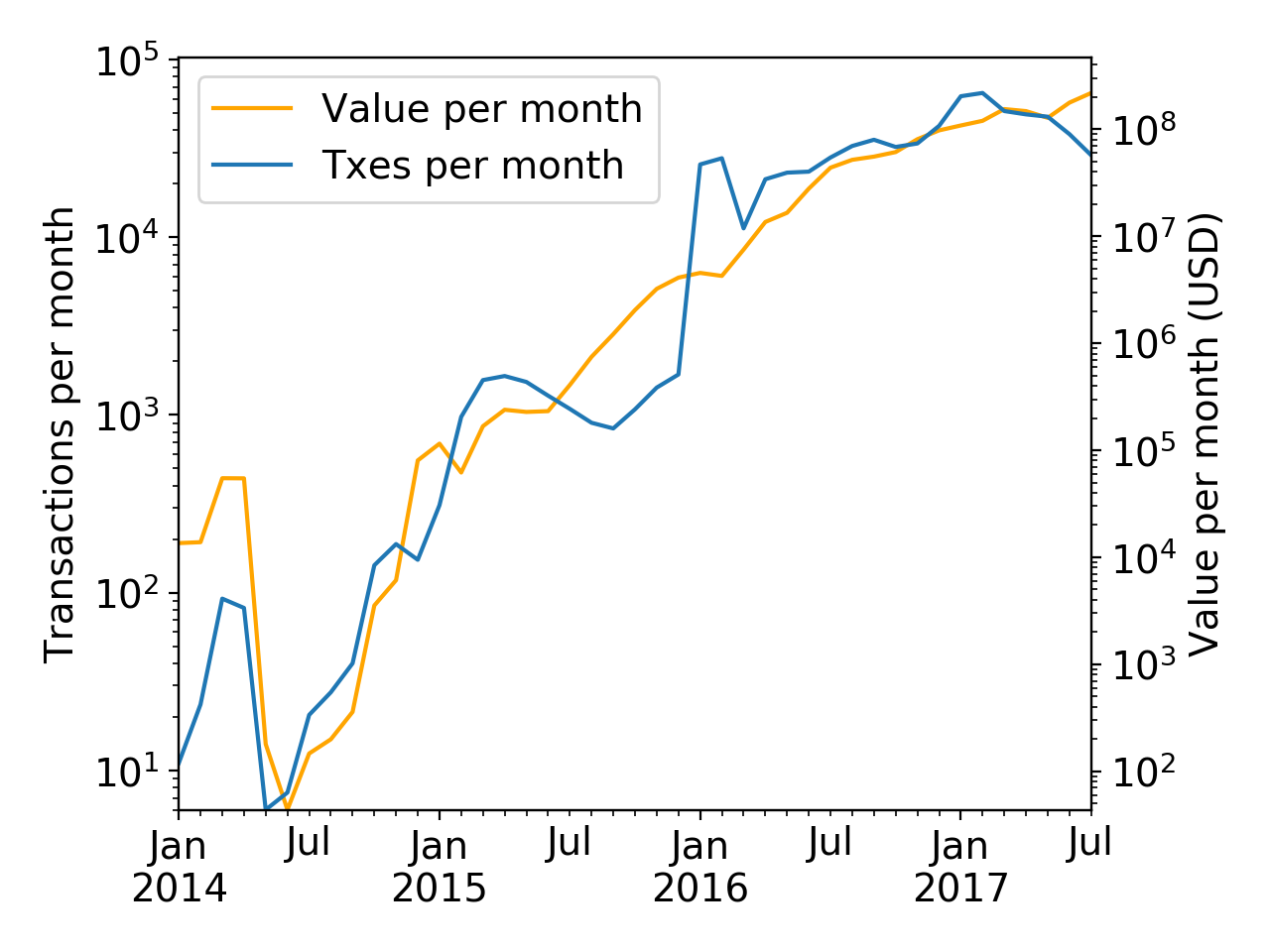}
\caption{Frequency and value of multisig transactions that expose confidential information about access structure changes on the blockchain.}\label{fig:multisig-confidentiality}
\end{figure}
    
{\bf Confidentiality.} 
For companies or individuals that use multisig to enforce access control over their wallet, multisig publicly exposes the access control structure as well as changes to that structure. In other words, it exposes the number of total keys and the number of keys needed to sign, as well as events that might trigger a change in access control such as a loss of a device or a departure of an employee.

Two characteristics indicate that a transaction might represent a change in access control:

\begin{itemize}
\item Single input, single output. Payment transactions typically involve multiple inputs and/or change outputs. By contrast, a transaction with only one input and one output (whether a regular or a multisig address) suggests that both are controlled by the same entity.
\item Overlapping sets of multisig keys between the input and the output, which suggests a change in access control but not a complete transfer of control.
\end{itemize}

As an example of such a transaction with these characteristics, consider the transaction 96d95e...\footnote{\href{https://blockchain.info/tx/96d95eb77ae1663ee6a6dbcebbbd4fc7d7e49d4784ffd9f5e1f3be6cd5f3a978}{\UrlFont https://blockchain.info/tx/96d95eb77ae1663ee6a6dbcebbbd4fc7d7e49d4784ffd9f5e1f3b\\e6cd5f3a978}}. In this transaction, over USD 130,000 of Bitcoin was transfered from one 2-of-3 multisig address to a second 2-of-3 multisig address. These addresses shared 2 keys in common, but one of the original keys was replaced with a different key.  Chainalysis\footnote{\url{https://www.chainalysis.com/}} labels both the input and output addresses as being controlled by \url{coinsbank.com}. This publicly reveals an internal restructuring happening at a private company.

In Figure \ref{fig:multisig-confidentiality} shows the total number and value of multisig transactions that publicly expose confidential access structure changes in this way. 

\begin{figure}
\centering
\includegraphics[width=\columnwidth]{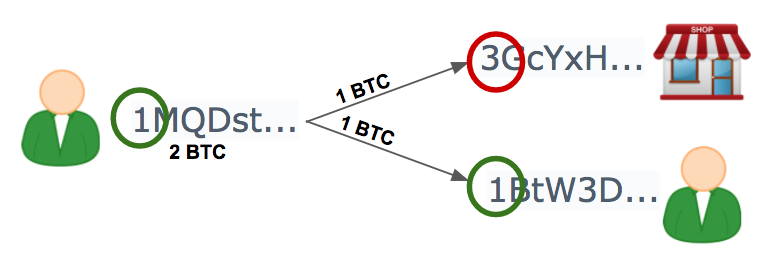}
\caption{A user pays a merchant that uses a multisignature (P2SH) address. It is easy to identify the change address because regular addresses look different from P2SH addresses.}\label{fig:multisig-anon}
\end{figure}

{\bf Privacy.}
As shown in Figure \ref{fig:multisig-anon}, the use of multisig provides a powerful heuristic for identifying the change address in a transaction. This is based on the intuition that a change address has the same access-control policy as the input address. We find that for many transactions, this heuristic allows identifying change addresses even though previously known heuristics~\cite{meiklejohn2013fistful} {\em don't} allow such a determination.

\begin{figure}
\centering
\includegraphics[width=\columnwidth]{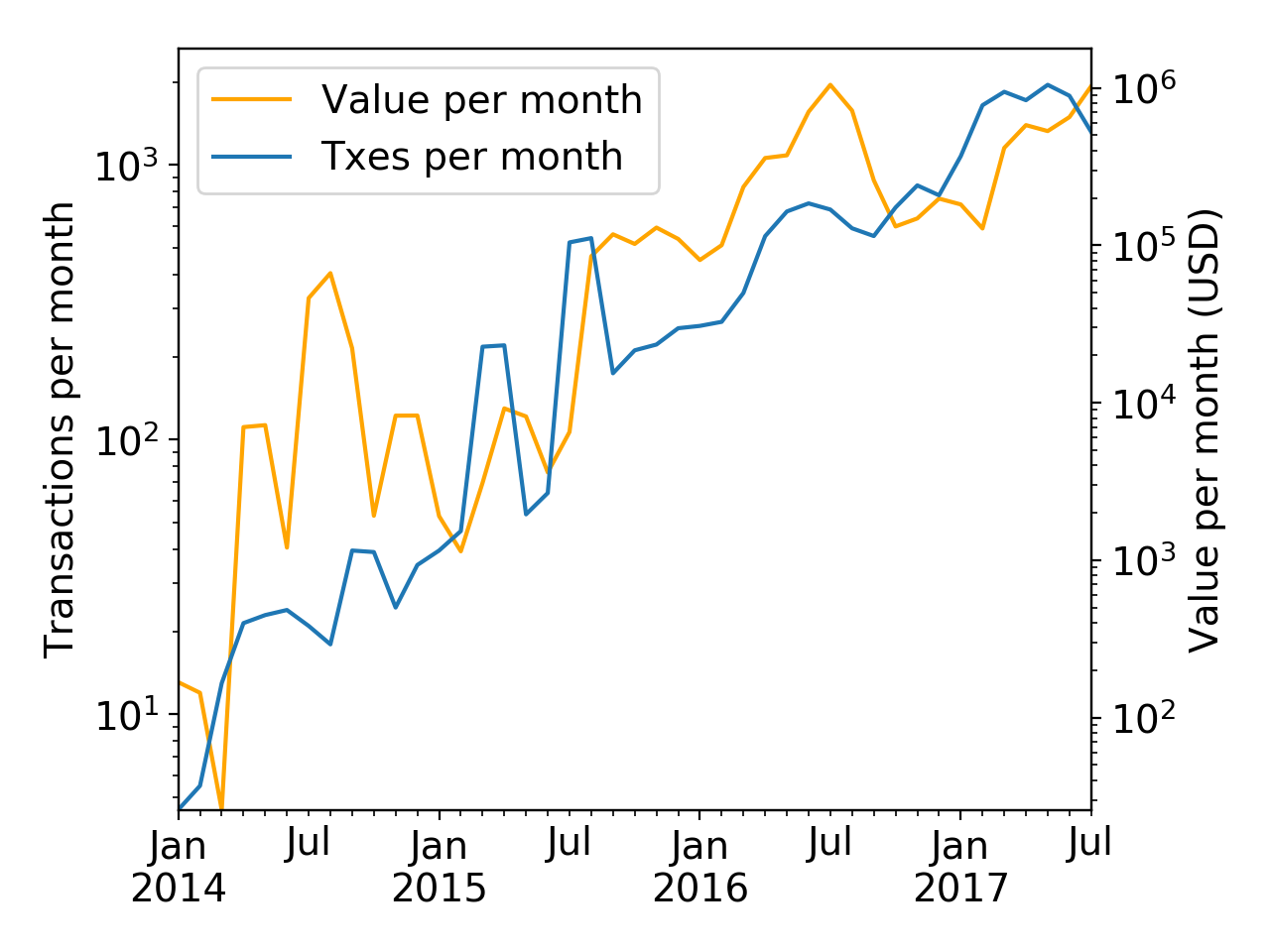}
\caption{Frequency and value of transactions that weaken multisig security by temporarily sending coins to regular addresses, advertising the presence of a single point of failure.}
\label{fig:multisig_security}
\end{figure}

\begin{figure*}
\centering
\includegraphics[width=2\columnwidth]{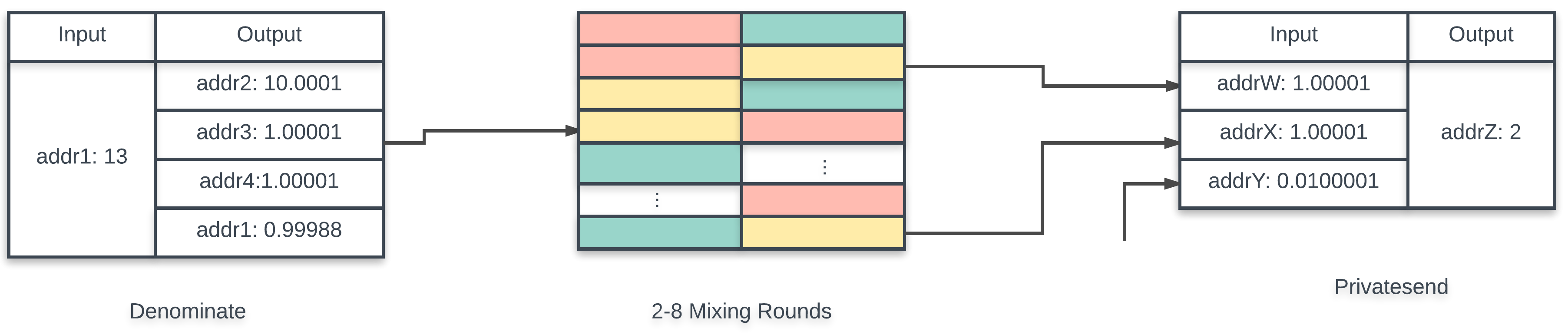}
\caption{{\bf Overview of Dash privacy.} First, in the Denominate step, a coin is broken down into valid denominations and the remainder is returned to the original address. Here, \texttt{addr2}, \texttt{addr3}, and \texttt{addr4} are the new denominated coins and the leftover 0.99988 Dash is sent back to \texttt{addr1}. Then for each denominated coin, there will be 2--8 rounds of mixing. When a user wishes to make a PrivateSend, the wallet will use these mixed coins as inputs. The input amount must be a multiple of the smallest denomination. Additionally another mixed input will be included as a fee. Here, the first two inputs provide the value for the output. The third input is for the fee. This value will generally be 0.0100001 Dash, but if coins of that denomination are not available, the wallet selects a mixed coin of the smallest denomination it possesses.
}\label{fig:dash-overview}
\end{figure*}

While Gennaro et al. mention the unfortunate privacy-infringing side-effect of multisig~\cite{ggn16}, we provide the first empirical evidence for the pervasiveness of this effect. Using BlockSci, we first applied previously known heuristics to every transaction in the blockchain, and found that they succeed in identifying 88,339,789 change addresses. We then augmented the change address detection by exploiting the privacy leaks of multisig, and  we were able to identify an additional 22,275,033 change addresses, an increase of over 25\%. Of the new change addresses that we identified, over  8 million were cases in which the anonymity of {\em non-multisig} users was weakened because they transacted with a party that used multisig (the scenario shown in Figure \ref{fig:multisig-anon}). Over 13 million were cases of multisig users weakening their own anonymity (i.e., the reverse scenario, in which a multisig user makes a payment to either a regular address or a multisig address with a different access structure.)

{\bf Security.}
A surprising, but relatively common motif is for multisig users to switch their money from a multisig address to a regular address, and then back into a multisig address. We conjecture that this may happen when users are changing the access control policy on their wallet, although it is unclear why they transfer their funds to a regular address in the interim, and not directly to the new multisig address. 

This practice negates some of the security benefits of multisignatures,  as it advertises to an attacker when a high-value wallet is most vulnerable. To identify this pattern, we looked for transactions in which all of the inputs were from multisig addresses of the same access structure and there was a single non-multisig output, which was subsequently sent back to a multisig address. We restricted our analysis to single output transactions as this is an indicator of self-churn --- i.e., a user shuffling money among her own addresses.

In Figure \ref{fig:multisig_security}, we show the number of transactions per month that exhibit this pattern of temporarily reducing security of a multisig address. We also show the total value of the outputs that were shuffled in this manner.

\subsection{Cluster intersection attack on Dash}
\label{sec:dash}

Goldfeder et al. recently showed the effectiveness of the cluster intersection attack against Bitcoin mixing \cite{cookie-meets-blockchain}. The attack seeks to link mixed coins to the cluster of wallet addresses that originally held the coins before mixing. The intuition behind the attack is that outputs mixed in different transactions are often spent together. Thus, when these coins are spent together, we trace each one back to a (potentially large) set of possible address clusters and examine the intersection of these sets. This will likely result in a unique cluster. We conclude that the mixed outputs are linked to the wallet represented by this cluster.

This is a significant weakness of mixing as an anonymity technique. In this section we provide evidence that Dash, a cryptocurrency designed with mixing in mind, is susceptible to this attack.

{\bf Overview of Dash.} Dash is one of three popular privacy-focused altcoins (alternative cryptocurrencies), along with Monero and Zcash. It is the largest of the three by market capitalization as of August 2017 --- over USD 2 billion. It is supported by a handful of vendors and a few alternative payment processors \cite{dash-use}. Dash is a fork of Bitcoin with a few key changes. It has a shorter block time (from 10 to 2.5 minutes) and uses the X11 hashing algorithm. It also has a two-tiered network, where nodes controlling 1,000 Dash or more have the option of becoming ``Masternodes'' --- full nodes that participate in the consensus algorithm, facilitate special types of transactions, and get a cut of the mining reward for their service. One of these special types of transactions is PrivateSend.
 
Dash's PrivateSend uses CoinJoin-style mixing, whereas Monero uses mixing based on ring signatures and Zcash provides cryptographic untraceability, which is a stronger (and provable) anonymity property. Mixing is not mandatory in Dash, but it is integrated into the default wallet and therefore easy to use. When a user chooses to start mixing, all her coins (up to a configurable limit with a large default value) are mixed with several rounds of mixing. The number of rounds is also configurable, but the default is 2. These mixed coins are then available for PrivateSend transactions.
 
Mix transactions in Dash use power-of-10 denominations. Therefore coins are broken up into these standard sizes before mixing is initiated. The mix transactions themselves each have three participants, each of whom contributes between 5 and 9 coins to be mixed. Finally, the PrivateSend transactions spend a set of mixed power-of-10 denominated outputs. Each of these three types of transactions has a distinct signature that is readily detectable on the Dash blockchain. In particular, the denominations are $1.00001 * 10^k$ instead of exactly $10^k$, and thus the values are highly unlikely to occur by chance. See Figure \ref{fig:dash-overview}.

\begin{figure}
\centering
\includegraphics[width=\columnwidth]{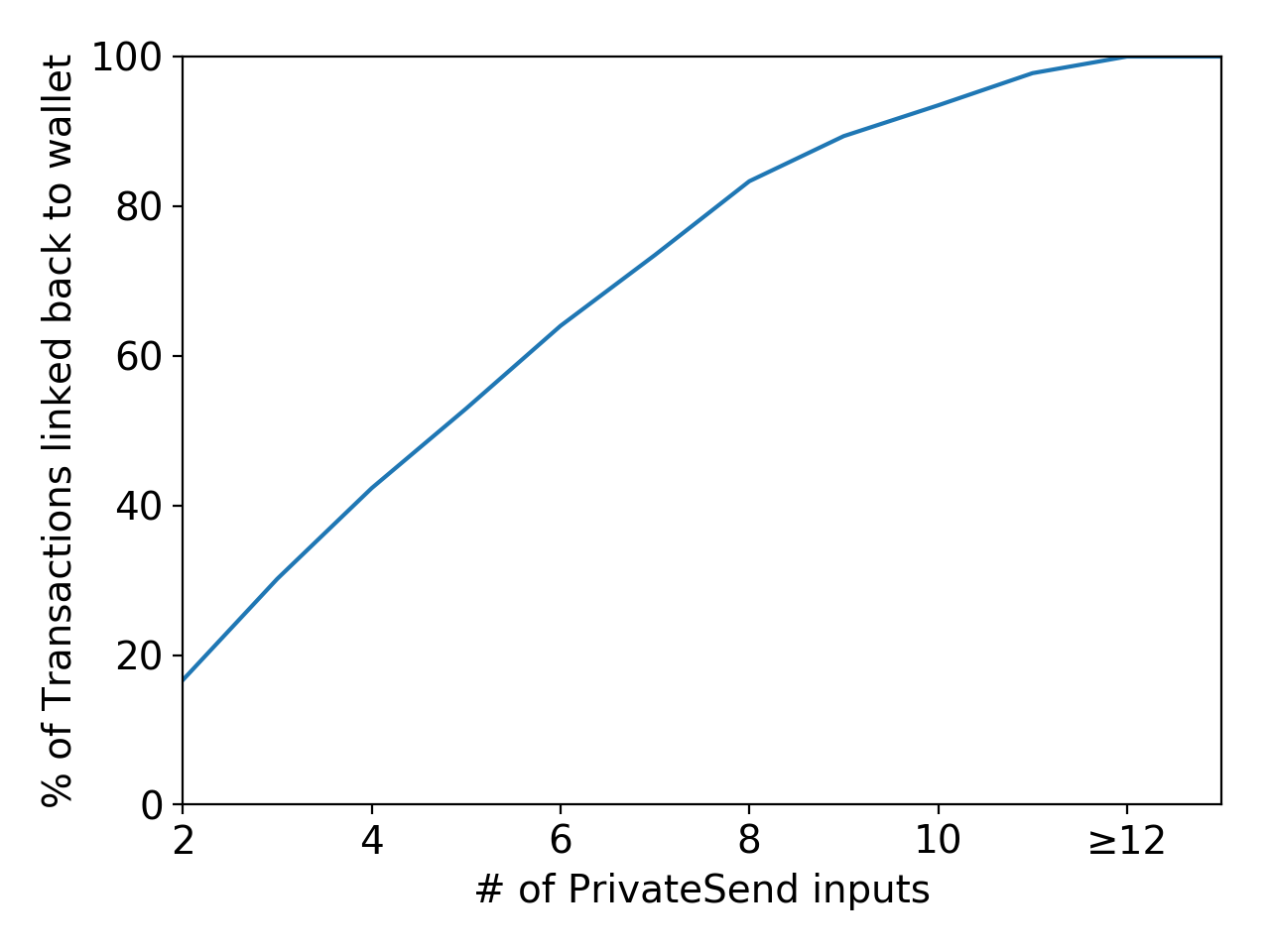}
\caption{Success rate of the cluster intersection attack on simulated Dash PrivateSend transactions as a function of the number of inputs.}\label{fig:dash-cluster-intersection}
\end{figure}

{\bf Dash and cluster intersection.}
Two features of the PrivateSend implementation combine to make Dash especially vulnerable to the cluster intersection attack. First, change addresses are not allowed for these transactions. This means that PrivateSend spenders must produce ``exact change'', which requires combining a large number of coins. Second, the denominations being powers of 10 (as opposed to, say, powers of 2) further increases the number of inputs in a typical transaction. For example, to pay 85 Dash, the sender must combine at least 8+5=13 inputs to avoid losing money. Figure \ref{fig:privatesend-input-count-distribution} in the Appendix shows the distribution of the number of inputs in PrivateSend transactions. Most such transactions have 3 or more inputs;  the mean is 40.1 and the median is 12.

Due to the large number of inputs, no auxiliary information is necessary to carry out the cluster intersection attack on Dash. The adversary --- anyone observing the public blockchain --- can infer that all inputs to a PrivateSend must trace back to the same wallet cluster. Thus, in the above example of a payment of 85 dash, the adversary knows that all 13 sets of clusters must have an element in common. The chance that there is more than one such cluster gets smaller and smaller as the number of clusters increases.

Of course, auxiliary information can make this attack more powerful. Beyond the risks posed by tracking cookies in \cite{cookie-meets-blockchain}, the Masternodes learn the input-output linkage for the mixing rounds that they facilitate. The privileged status of Masternodes in the Dash p2p network raises other potential privacy vulnerabilities \cite{WarningD62:online}, but that is not our focus.

{\bf Experimental setup.}
To perform this attack, we used shapeshift.io (an online service for conversion between cryptocurrencies) to convert Bitcoin into Dash, which we withdrew into a single address. We used the default Dash wallet to mix 0.55 Dash using the default parameters, namely 2 rounds of mixing. We obtained 55 separate mixed outputs, each 0.01 Dash.
 
Next, we re-implemented the PrivateSend algorithm from the Dash wallet code on top of BlockSci. Given a desired spend amount, the algorithm selects a set of mixed inputs from the wallet that sum to this amount. It is shown in Algorithm \ref{alg:walletsim} in the appendix. This allowed us to simulate our own PrivateSend transactions instead of actually making them. The latter would have required paying a transaction fee for each data point; generating the data shown below would have required spending several hundred USD worth of Dash in transaction fees, and holding several tens of thousands of USD worth of Dash.

For each of the simulated PrivateSends, we ran the cluster intersection attack. We consider the attack successful if it results in a unique cluster of addresses, namely the single address that we started from.

{\bf Results.} Figure \ref{fig:dash-cluster-intersection} shows the success rate of the cluster intersection attack, showing a sharp increase in accuracy as the number of inputs increases. For transactions with 12 or more inputs (coincidentally, the median number of inputs of PrivateSend transactions on the blockchain), the attack is always accurate.

In the above experimental setup, we started from a single pre-mixing address holding Dash. In reality, users may obtain Dash in multiple installments and hold these coins in their wallet in a manner that is not easily linkable to each other. Relying on this is unwise for privacy, as it is a form of security through obscurity; nevertheless, it is a factor that will significantly hurt the accuracy of the attack in practice. Evaluating the attack on existing PrivateSend transactions is challenging due to the lack of ground truth, and is a topic for future work.

\subsection{The block space market}
\label{sec:blockspace}

\begin{figure}
\centering
\includegraphics[width=\columnwidth]{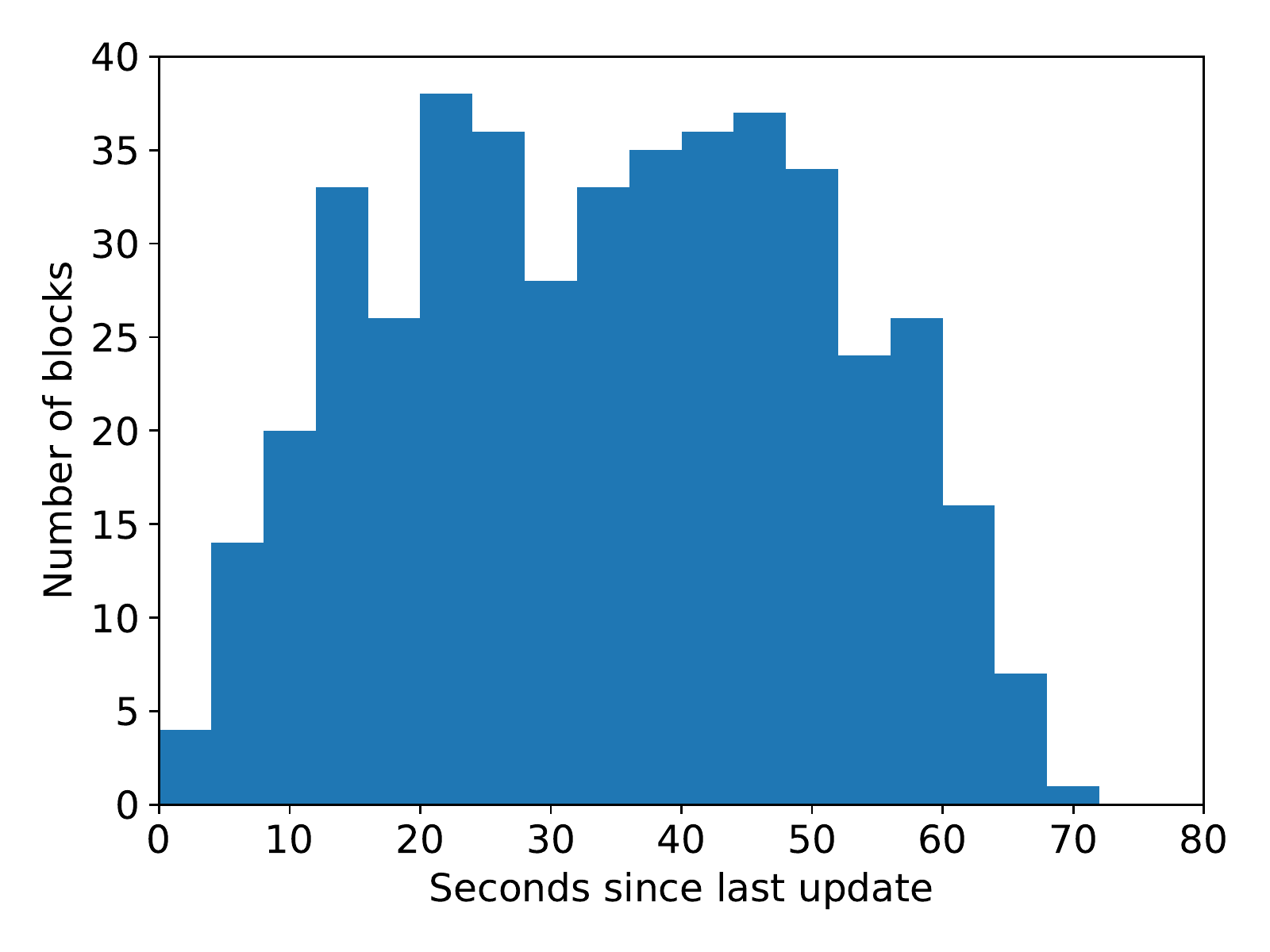}
\caption{Distribution of the apparent gap between the most recent transaction in a block and the block time for Antpool over a 2-week period in July 2017, suggestive of a 60-second block update interval.}\label{fig:antpool}
\end{figure}

Blockchains are massively replicated, and so most blockchain protocols limit the size of blocks. In Bitcoin, the limit is currently 1MB per block, which translates to a few thousand transactions per ten-minute interval. The demand for inclusion in the blockchain exceeds this rate, and thus a market for block space has developed.\footnote{Bitcoin Cash, on the other hand, appears committed to indefinitely increasing the block size limit to keep pace with demand.} A rational miner should include a set of transactions that maximizes the revenue of the block, roughly equivalent to filling blocks with transactions in decreasing order of transaction fee per byte.

In this analysis, we examine two ways in which miners depart from this simple revenue-maximizing transaction-selection algorithm. Neither appears to have been widely discussed. 

{\bf Slow block updates.} Mining involves two steps: creating valid blocks by assembling transactions, and computing the block hash with different values for the `nonce' field. It is only the second step that is computationally intensive, and has been the source of much innovation in mining hardware and business models. The first step is computationally trivial. However, to maximize revenue, it must be repeated as soon as a new transaction arrives. If a miner instead updates their
blocks, say, once a minute, they will leave transaction fees on the table.

To test if any miners or mining pools have slow block update times, we compute (for each block) the time delay between the most recent transaction that was included in the block and when the block itself was broadcast. Network latency adds some error to our estimates of when the pool saw a transaction and when the miner computed the block, but these are on the order of a few seconds, much smaller than the lag we are interested in. In particular, we find that for Antpool's blocks, the time delay is roughly uniformly distributed between 0 and 60 seconds, consistent with a conjecture that Antpool's blocks are updated every 60 seconds (Figure \ref{fig:antpool}).

\begin{figure}
\centering
\includegraphics[width=\columnwidth]{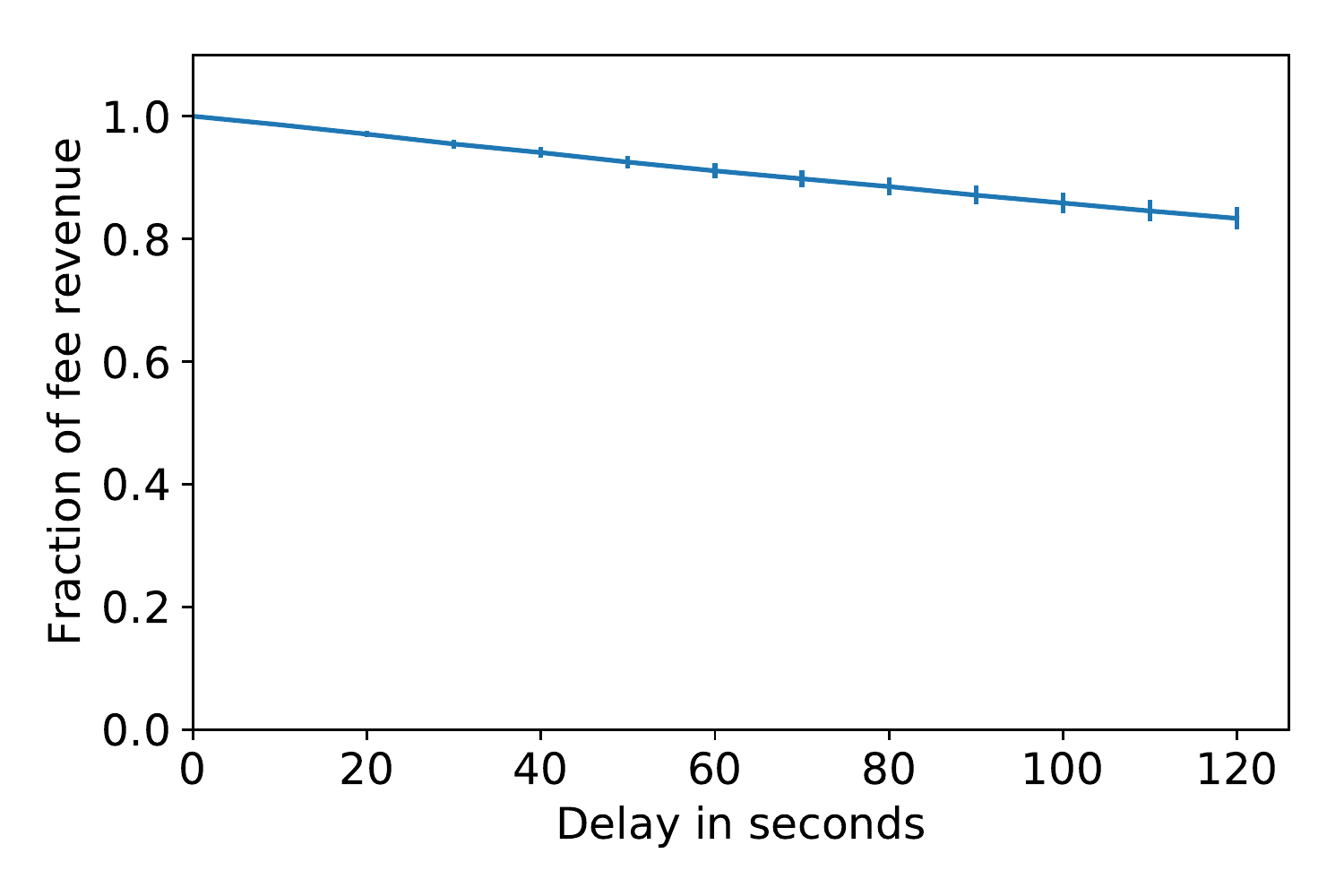}
\caption{Relative loss in transaction fee due to slow block updates. Error bars represent 95\% confidence intervals.}\label{fig:delay-vs-loss}
\end{figure}

In Figure \ref{fig:delay-vs-loss} we show how much a miner or pool would lose in transaction fees (assuming the transaction fee distribution seen in late July 2017) for various values of the block update interval. A miner with a 60-second interval would lose an average of 5\% of transaction-fee revenue in each block. Given Antpool's share of hashpower and the exchange rate of BTC, we estimate that Antpool (and its participants) lost up to USD 90,000 compared to a hypothetical scenario in which there is no delay in updating blocks.

Appendix \ref{sec:slow-block} contains similar figures for other top mining pools. In most cases the block updates appear to be surprisingly slow. One explanation is that the default values of the update interval for Stratum and other mining protocols appears to be 60 seconds \cite{pool-update-rate}. 

With modern mining protocols such as Stratum, miners need to download only the block headers, and not the contents of blocks, from pool operators, so a much smaller value of the update interval should be feasible.

\begin{figure}
\centering
\includegraphics[width=\columnwidth]{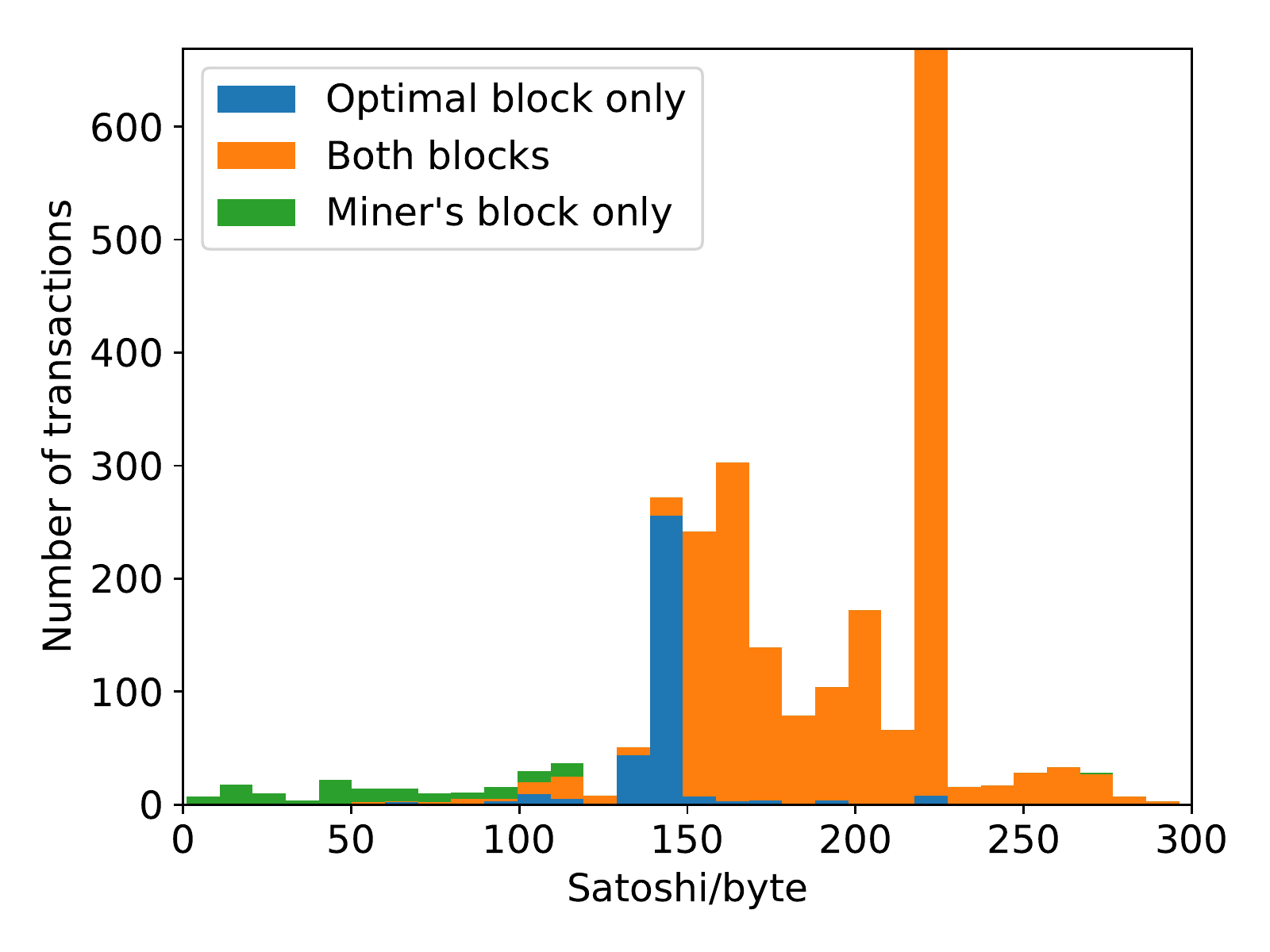}
\caption{Example of a BW.COM block (at height 478458) that includes multiple low-fee transactions.}\label{fig:bw}
\end{figure}

{\bf Unexplained inclusion of low-fee transactions.} We tested if there are miners that claim less than the available transaction fees for reasons {\em other} than slow block update times. To test if a block $B$ is in this category, we create a valid block out of the available transactions in the mempool at the time of the most recent transaction found in $B$. The most interesting miner showing this behavior is BW. We find that 12\% of BW's blocks contain transactions that pay far less transaction fees than the minimum that would be necessary for inclusion by a revenue-maximizing miner. Figure \ref{fig:bw} illustrates such a block. 

We examined a set of 2,148 such suspiciously low-fee transactions in July 2017. We defined these as transactions paying significantly lower fees (5 satoshis/byte or more) than the lowest fee in the optimal block. When constructing the optimal block we accounted for the possibility of ``child pays for parent" transactions, and excluded them from the set of suspiciously low-fee transactions. These are low-fee transactions that may get included if a child transaction (i.e. a transaction that spends one of its outputs) pays high enough fees.
We observed several patterns that could potentially explain the  inclusion of low-fee transactions (summarized in Table \ref{tab:low-fee-results}).

\begin{itemize}
\item {\em Priority.} Bitcoin Core has a notion of `priority' of transactions that includes factors such as the age of the coins being spent. About a third of the low-fee transactions had a high priority score, which is potentially why they were included. Priority is a vestige of the era before the emergence of the block space market, and aimed to disincentivize transactions that wasted block space; we are aware of no good reason for miners to reserve space in blocks for them. Yet this practice appears to persist.
\item {\em Zero-fee transactions not previously seen.} About 29\,\% of transactions paid no fee, and were transactions that we had not recorded before their appearance in the block. This suggests private relationships between the creators of those transactions and the miners who include them. Coinbase is one such entity whose transactions are not always broadcast publicly.
\item {\em Sweep transactions}. About a fifth of transactions were ``sweep'' transactions with over 10 inputs and only one output. It is not clear why miners would include them despite their low fee per byte; perhaps miners reserve some space for transactions with sufficiently high absolute fees.
\end{itemize}

\begin{table}
\centering
\begin{tabular}{l r}
\toprule
Characteristic & \# transactions \\ \midrule
High priority &   770 \\
Zero fee &        634 \\
Sweep  &  411 \\ 
Unexplained &     802 \\ \midrule
Total &     2,148 \\
\bottomrule
\end{tabular}
\caption{Characteristics of low-fee transactions that may explain their inclusion in blocks (not mutually exclusive).}\label{tab:low-fee-results}
\end{table}

This still leaves about 37\% of transactions unexplained by any of the above patterns.

Overall, the inclusion of these low-fee transactions cost miners and mining pools about 20 bitcoins during the two-week period of observation, roughly equivalent to over USD 100,000 in July 2017. BW.com, in particular, lost 0.065 bitcoins {\em per block}, equivalent to a few hundreds of dollars.

At the time of writing, transaction fees remain a small fraction (about 20\%) of the total mining reward, but the losses due to sub-optimal block construction will gain in importance if transaction fees continue to increase.

\subsection{Improved estimates of the velocity of cryptocurrencies}
\label{sec:velocity}
The velocity of money is the frequency with which one unit of currency is used for purchases in a unit of time. It can provide an insight into the the extent to which money is used as a medium of exchange versus a store of value.

In most cases it is not possible to infer the purpose behind a cryptocurrency transaction from the blockchain. However, an alternative definition of the velocity of money is the frequency with which one unit of currency changes possession in any manner (whether or not for purchases of goods and services) in a unit of time. Blockchain analysis may enable estimating the velocity of cryptocurrencies under this definition. 

\begin{figure}
\centering
\includegraphics[width=\columnwidth]{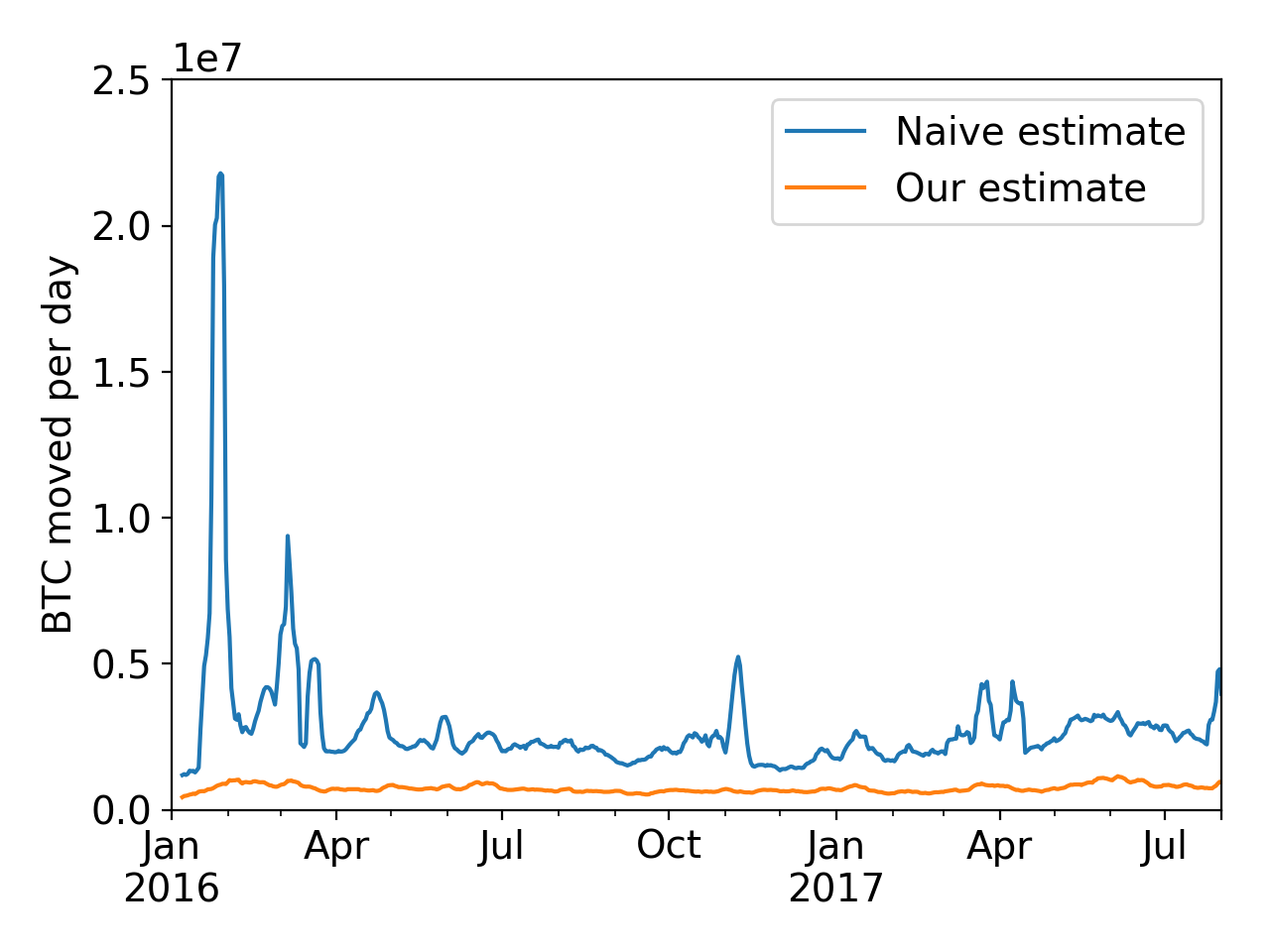}
\caption{Two estimates of the velocity of bitcoins.}\label{fig:total-daily-output-value}
\end{figure}

Even under this simplified definition, it is challenging to estimate the velocity of cryptocurrencies. A naive method would be to compute the total value of transaction outputs in a unit of time and divide it by the total value of the money supply during that period. However, multiple addresses may be controlled by the same entity, and therefore not all transaction outputs represent changes in possession. Meiklejohn et al. call this ``self churn'' \cite{meiklejohn2013fistful}, a term that we adopt. The impact of self churn is visually obvious in the graph of total transaction outputs (Figure \ref{fig:total-daily-output-value}). We would not expect spikes such as those on January 27, 2016 and April 23, 2017 if the graph reflected actual money demand, which would be much more stable over time. 

To minimize the effect of self churn, we adopt two heuristics. First, we eliminate outputs controlled by an address linked to one of the inputs addresses (as defined in Section \ref{sec:analysis-library}, but after removing the ``supercluster" to minimize false positives). This eliminates change outputs, and entirely eliminates transactions that are detectable as an entity ``shuffling their money around''. We also eliminate outputs that are spent within less than $k$ blocks (we use $k=4$). Manual examination suggests that such transactions are highly likely to represent self-churn, such as ``peeling chains'' where a large output is broken down into a series of smaller outputs in a sequence of transactions.

The orange line in Figure \ref{fig:total-daily-output-value} shows the daily Bitcoin transaction volume after applying the above two heuristics. With this estimate, the velocity of Bitcoin works out to 1.4 per month averaged over the period January 2016--July 2017, compared to 5.4 with the naive metric. Our revised estimate is not only much lower but also much more stable over time.

We note several caveats. First, this still likely fails to exclude some transfers of value between addresses controlled by the same entity. Without ground truth, it is hard to be certain how good the estimate is. Second, it doesn't count transfers of possession that don't touch the blockchain. When exchanges, online wallets, and other intermediaries hold money on behalf of users, payments and transfers of ``bitcoins" might happen even though no actual bitcoins changed hands. Nevertheless, we believe that the metric can be a useful proxy for understanding the use of cryptocurrencies, and possibly for comparing between cryptocurrencies. 

\begin{figure}
\centering
\includegraphics[width=\columnwidth]{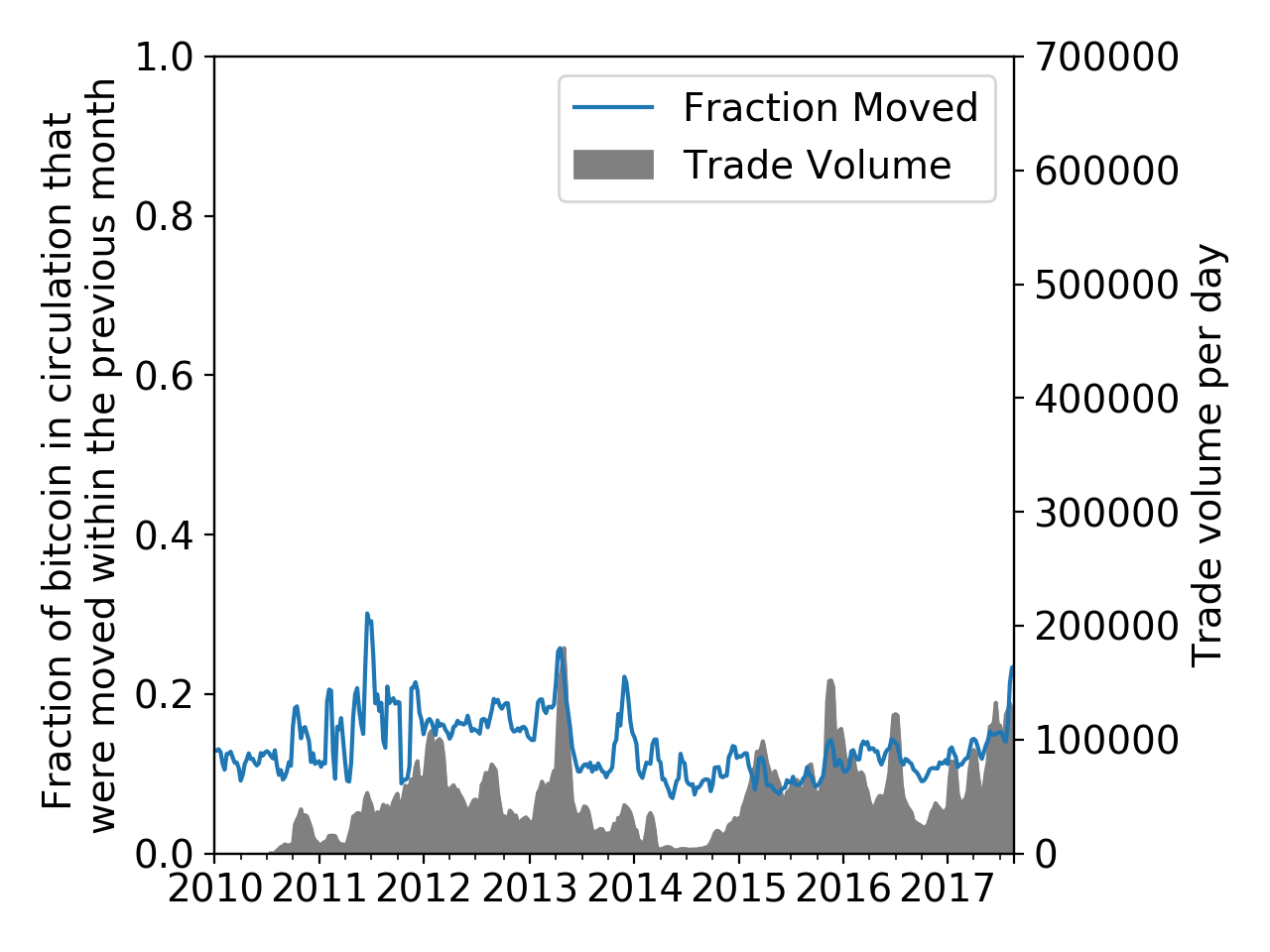}
\caption{The fraction of bitcoins moved in the previous month, and the USD-BTC trade volume.}\label{fig:medium-vs-store}
\end{figure}

Another measurement that may help distinguish between bitcoins used as a medium of exchange and as a store of value is shown in Figure \ref{fig:medium-vs-store}. At any given time, most transaction outputs (on average, 86\%) have been sitting unspent for over a month. The high values pre-2013 are attributable to the gambling service SatoshiDice, an observation also made by Meiklejohn et al. \cite{meiklejohn2013fistful} (the drop in May 2013  coincides with SatoshiDice blocking U.S. players). Superimposing the BTC-USD trade volume shows that many of the spikes (e.g. April 28, 2013, April 2, 2017) correspond to speculative bubbles. Overall, the graph suggests that only a small percentage of bitcoins are used for activities other than investment and speculation, although that fraction has been gradually increasing over the past year.

\section{Conclusion} There is a high level of interest in blockchain analysis among developers, researchers, and students, leading to an unmet need for effective analysis tools. While general-purpose in-memory graph databases exist \cite{dubey2016weaver}, a tool customized to blockchain data can take advantage of its append-only nature as well as provide integrated high-performance routines for common tasks such as address linking.

BlockSci has already been in use at Princeton as a research and educational tool. We hope it will be broadly useful, and plan to maintain it as open-source software.

\begin{acks}
We are grateful to Lucas Mayer for prototype code, Jason Anastasopoulos, Sarah Meiklejohn, and Dillon Reisman for useful discussions, and Chainalysis for providing access to their Reactor tool. This work is supported by NSF grants CNS-1421689 and CNS-1651938 and an NSF Graduate Research Fellowship under grant number DGE-1148900.
\end{acks}

\bibliographystyle{IEEEtran}
\bibliography{blocksci}

\clearpage

\appendix

\section{Dash PrivateSend algorithm} 
\begin{algorithm}
\caption{PrivateSend wallet simulation. \\
\ \ \ Input: desired amount to spend in a PrivateSend\\
\ \ \ Output: a set of unspent outputs to add up to this value}\label{alg:walletsim}
\begin{algorithmic}[1]
\Procedure{SelectPSInputs}{$send\_amount$}
\State $\textit{T} \gets \text{set of transactions that have at least one }$
\par \hskip\algorithmicindent $\text{output that is unspent and owned by us}$
\State $\textit{T} \gets \text{Sort $T$ by (denomination, transaction hash)}$
\State $selected \gets \textit{\{\}}$
\For{\textbf{each} $t \in T$}:
\For{\textbf{each} $output \in t.outputs$}:
\If{$\Call{value}{selected}+\Call{value}{output} $ 
\\ \par \hskip\algorithmicindent \hskip\algorithmicindent \hskip\algorithmicindent \qquad $ > send\_amount$}
\State \textbf{break}
\EndIf
\State  $selected.insert(output)$
\If {$\Call{value}{selected} == send\_amount$}
\State \Return $selected$
\EndIf
\EndFor
\EndFor
\State\Return "Insufficient Funds"
\EndProcedure
\end{algorithmic}
\end{algorithm}

\section{Additional Figures}\label{sec:slow-block}

\begin{figure}[h]
\centering
\includegraphics[width=\columnwidth]{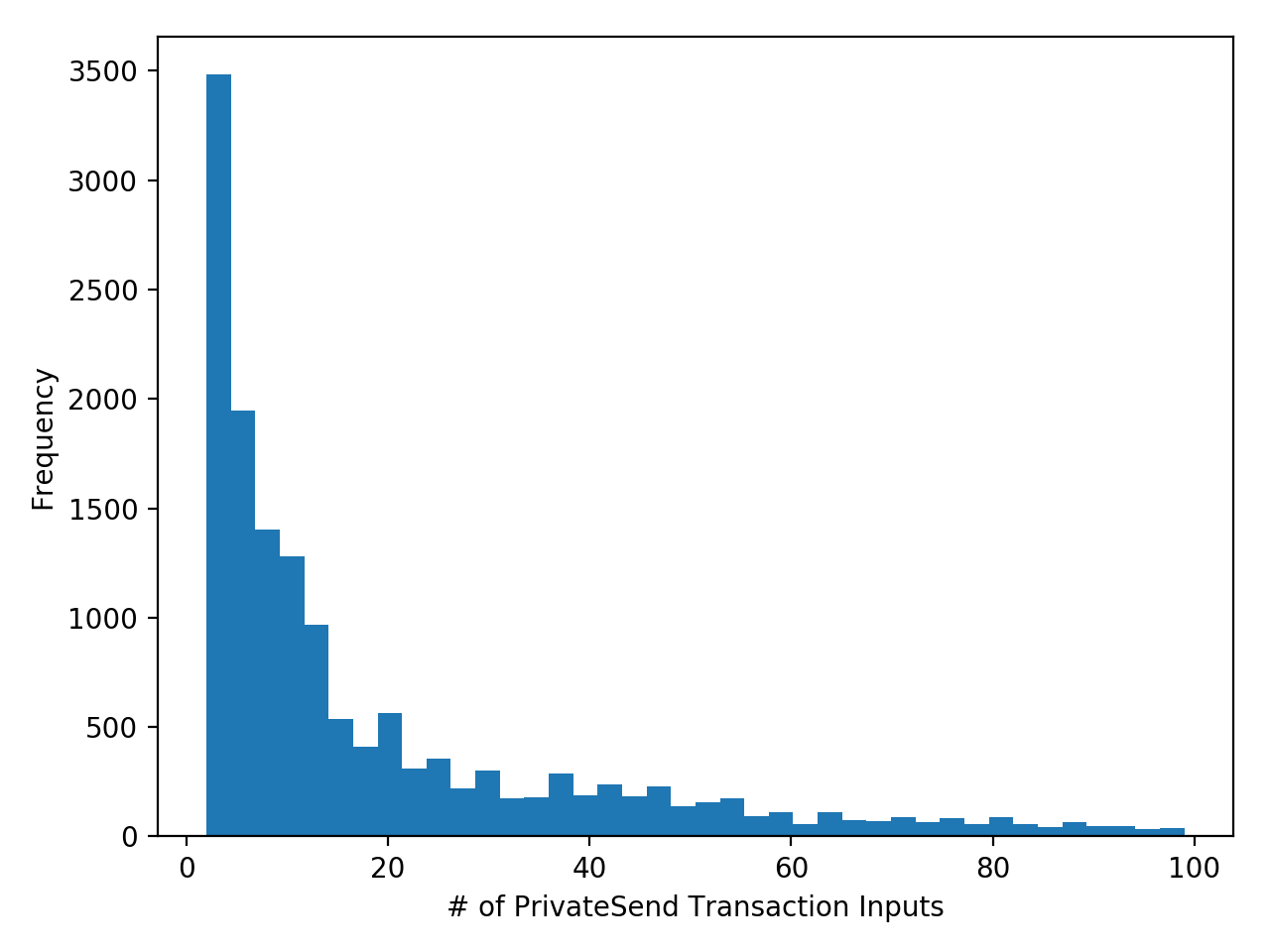}
\caption{Distribution of the number of inputs of Dash PrivateSend transactions}\label{fig:privatesend-input-count-distribution}
\end{figure}

\begin{figure}[h]
\centering
\includegraphics[width=\columnwidth]{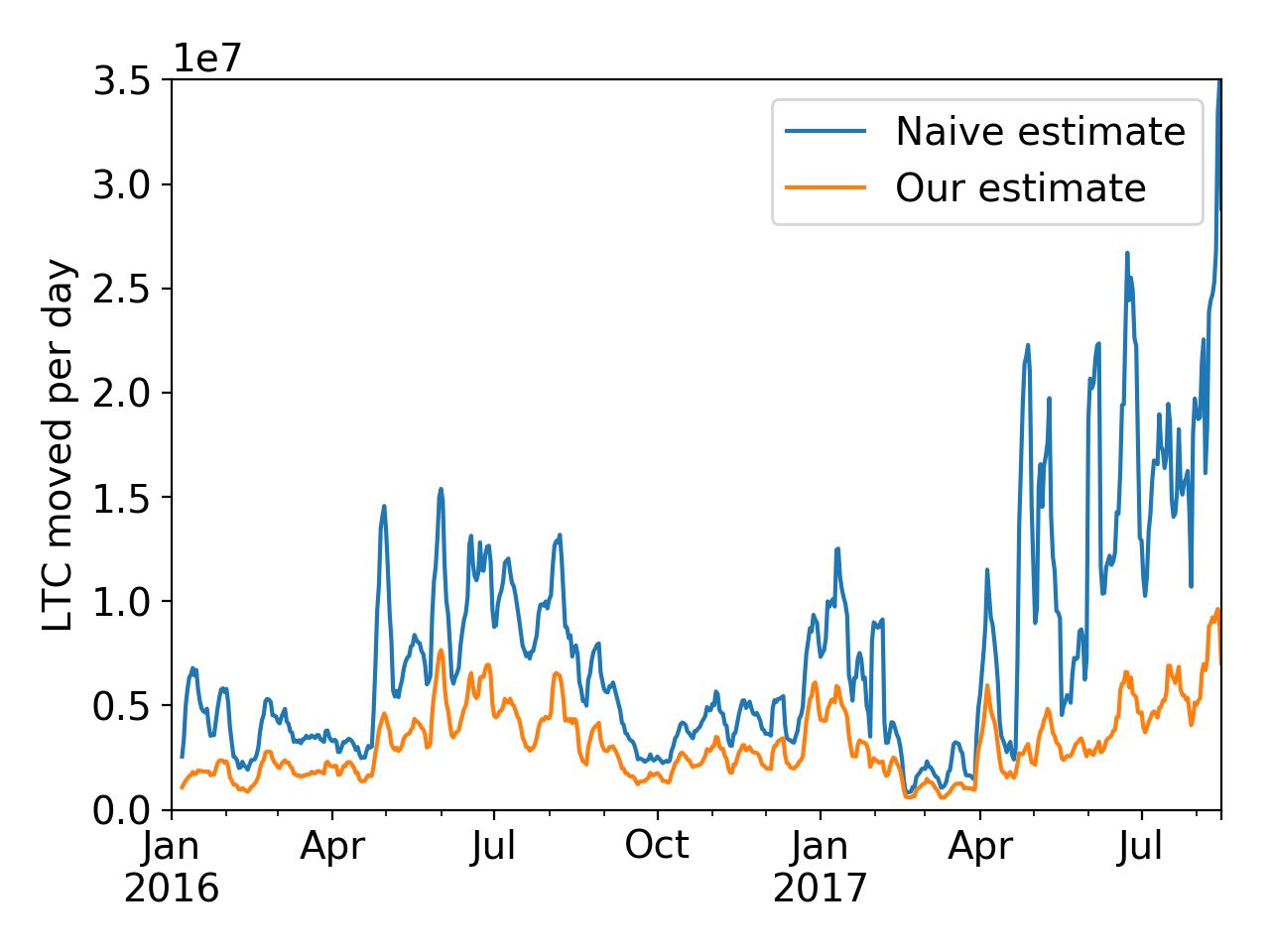}
\caption{Two estimates of the velocity of litecoins.}\label{fig:total-daily-output-value-altcoin}
\end{figure}

\begin{figure}[h]
\centering
\includegraphics[width=\columnwidth]{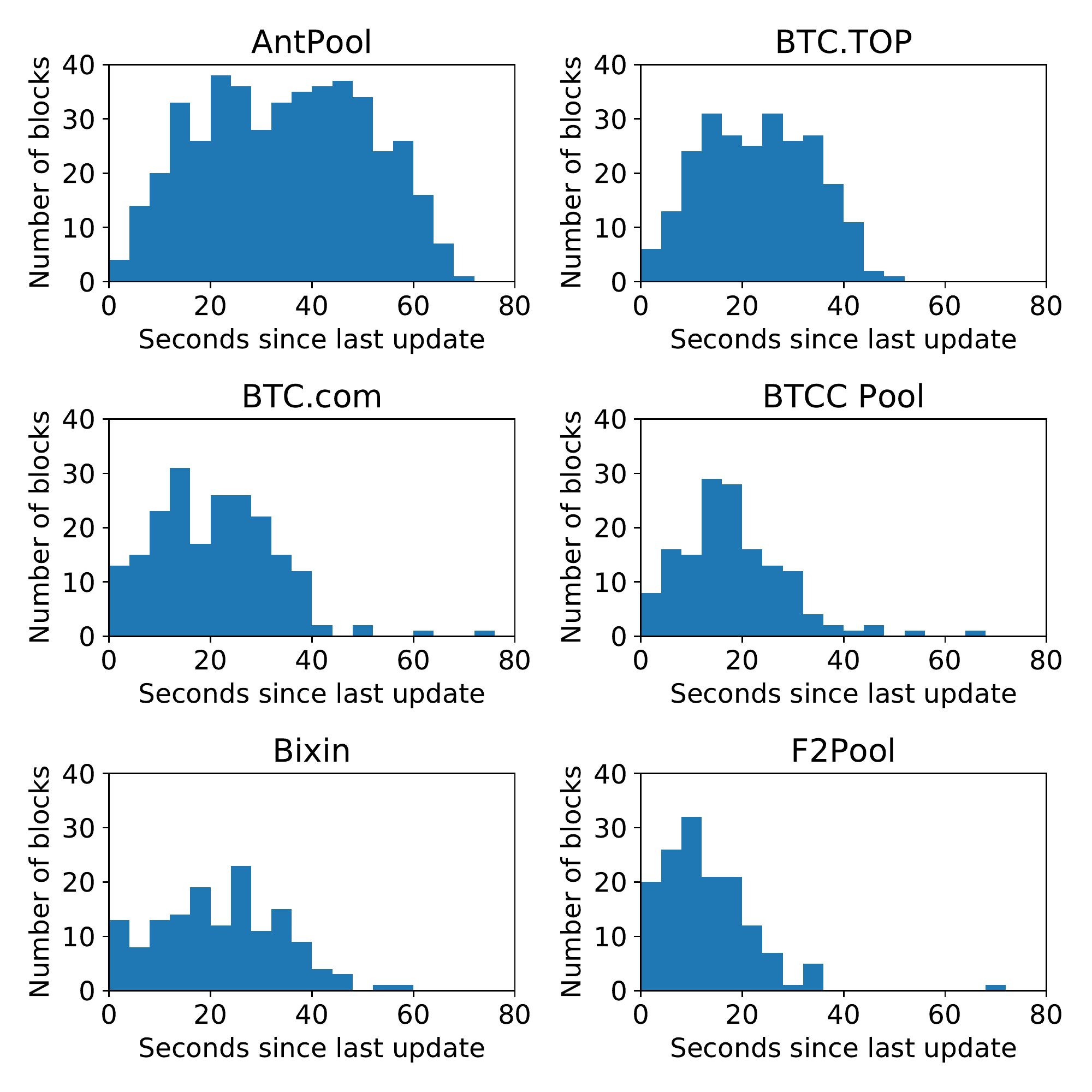}
\caption{Distribution of the apparent gap between the most recent transaction in a block and the block time, for the top 6 mining pools.}\label{fig:other-pools-delay}
\end{figure}

\addtolength{\textheight}{-12cm} 




\end{document}